\newif\ifsingle
\ifsingle
\documentclass[11pt,draftclsnofoot, onecolumn]{IEEEtran}	
\else		
\documentclass[10pt,final, twocolumn]{IEEEtran}
\fi
  
\usepackage{BKN}

\usepackage[all=normal,paragraphs=normal,floats=normal,mathspacing=normal,wordspacing=normal,charwidths=tight,mathdisplays=normal,leading=normal]{savetrees}
\definecolor{NewColor}{rgb}{0,0,0}%{0.2,0,0.5}
\definecolor{NewColor2}{rgb}{0,0,0}%{0.2,0,0.5}
\title{Bayesian KalmanNet: Quantifying Uncertainty in Deep Learning Augmented Kalman Filter}
\author{Yehonatan Dahan,~\IEEEmembership{Student  Member,~IEEE}, Guy Revach,~\IEEEmembership{Senior Member,~IEEE}, Jindrich Dunik,~\IEEEmembership{Senior Member,~IEEE}, and Nir Shlezinger,~\IEEEmembership{Senior Member,~IEEE}
\thanks{
Parts of this work were presented at the IEEE International Conference on Acoustics, Speech, and Signal Processing (ICASSP) 2024 as \cite{dahan2024uncertainty}. %\\
Y. Dahan and N. Shlezinger are with the School of ECE, Ben-Gurion University of the Negev,  Israel (e-mail: yehonatd@post.bgu.ac.il, nirshl@bgu.ac.il). %\\
G. Revach is with the Institute for Signal and Information Processing, %\\
D-ITET, ETH Zürich, Switzerland (email: grevach@ethz.ch). %\\
J. Dunik is with the Faculty of Applied Science, University of West Bohemia in Pilsen, Czech Republic (email: dunikj@kky.zcu.cz). J. Dunik has been partially supported by the Ministry of Education, Youth and Sports of the Czech Republic under project ROBOPROX - Robotics and Advanced Industrial Production CZ.02.01.01/00/22\_008/0004590.
}} 
\begin{document}
\maketitle
%
%%%%%%%%%%%%%%%%
%%% Abstract %%%
%%%%%%%%%%%%%%%%
%
\begin{abstract}
Recent years have witnessed a growing interest in tracking algorithms that augment \acp{kf} with \acp{dnn}. By transforming \acp{kf} into trainable deep learning models, one can learn from data to reliably track a latent state in complex and partially known dynamics. However, unlike classic \acp{kf}, conventional \ac{dnn}-based systems do not naturally provide an uncertainty measure, such as error covariance, alongside their estimates, which is crucial in various applications that rely on \ac{kf}-type tracking.
This work bridges this gap by studying error covariance extraction in \ac{dnn}-aided \acp{kf}. We begin by characterizing how uncertainty can be extracted from existing \ac{dnn}-aided algorithms and distinguishing between approaches by their ability to associate internal features with meaningful \ac{kf} quantities, such as the \acl{kg} and prior covariance. We then identify that uncertainty extraction from existing architectures necessitates additional domain knowledge not required for state estimation.
Based on this insight, we propose {\em Bayesian KalmanNet}, a novel \ac{dnn}-aided \ac{kf} that integrates Bayesian deep learning techniques with the recently proposed KalmanNet and transforms the \ac{kf} into a {\em stochastic} machine learning architecture. This architecture employs sampling techniques to predict error covariance reliably without requiring additional domain knowledge, while retaining KalmanNet's ability to accurately track in partially known dynamics. Our numerical study demonstrates that Bayesian KalmanNet provides accurate and reliable tracking in various scenarios representing partially known dynamic systems.
\end{abstract} 

\acresetall
%
%%%%%%%%%%%%%%%%%%%%
%%% Introduction %%%
%%%%%%%%%%%%%%%%%%%%
%
\section{Introduction}
Tracking the latent state of a dynamic system is one of the most common tasks in signal processing~\cite{durbin2012time}. Often, one is required not only to track the state, but also to provide a reliable assessment of the uncertainty in its estimate. For instance, navigation systems must not only provide the instantaneous position but also characterize its accuracy~\cite{bar2004estimation}. 
\textcolor{NewColor}{Moreover, uncertainty quantification in state estimation provides a measure of confidence in the predictions and enables informed, risk-aware decision-making, which is particularly critical in safety-sensitive domains.}
Uncertainty quantification is inherently provided by classic algorithms, such as the \ac{kf} and the \ac{ekf}, that rely on the ability to faithfully describe the dynamics as a closed-form \ac{ss} model. Such model-based algorithms track not only the state, but also compute the error covariance~\cite{gannot2008kalman}.  
%
%When operating with partially known \ac{ss} models, the conventional approach imposes a parametric model on the dynamics and aims to estimate it from data via system identification. This limits 
%

The emergence of deep learning gave rise to a growing interest in using \acp{dnn}  to carry out  tasks involving time sequences~\cite{lim2021time}. 
Various \ac{dnn} architectures can learn from data to track dynamic systems without relying on full knowledge of an underlying \ac{ss} model. These include generic \acp{rnn}~\cite[Ch. 10]{goodfellow2016deep} and attention mechanisms~\cite{vaswani2017attention}, as well as \ac{rnn}-type architectures inspired by \ac{ss} models~\cite{gu2021combining} and \ac{kf} processing~\cite{becker2019recurrent}. Such \acp{dnn} are trained  end-to-end   without using any statistical modeling.

When one has partial knowledge of the underlying dynamics, it can be leveraged by hybrid model-based/data-driven designs~\cite{shlezinger2020model, shlezinger2022model,shlezinger2023model}. Existing approaches to exploit partial domain knowledge can be roughly divided following the machine learning paradigms of {\em generative} and {\em discriminative} learning~\cite{shlezinger2022discriminative}. Generative approaches are {\em \ac{ss} model-oriented}, using \acp{dnn} to learn the missing aspects of the underlying \ac{ss} model, i.e., for system identification, while adopting parameterizations varying from physics-based \ac{ss} models~\cite{imbiriba2022hybrid,imbiriba2023augmented} to fully \ac{dnn}-based ones~\cite{chiuso2019system, gedon2021deep,ljung2020}. The learned models are then used by classical algorithms, such as the \ac{ekf}, thus typically preserving the limitations of these methods to, e.g., Gaussian temporally independent noises~\cite{shlezinger2024ai}. 

The discriminative strategy is {\em task-oriented}, learning to directly output the state estimate, following the conventional usage of \acp{dnn} for classification and regression~\cite{goodfellow2016deep}.  
Existing designs vary in the knowledge they require on the underlying model.  For settings with relatively high level of domain knowledge, e.g., when one can approximately characterize the dynamics as a linear Gaussian \ac{ss} model, \acp{dnn} can enhance classic \acp{kf} by providing correction terms~\cite{satorras2019combining} or pre-processing~\cite{he2014state} trained alongside the tracking algorithm. In scenarios with only partially known \ac{ss} models, \ac{dnn} augmented \acp{kf} were proposed, converting the \ac{kf} into a machine learning algorithm~\cite{revach2022kalmannet,choi2023split,ni2022rtsnet,song2024practical, wang2024nonlinear,ghosh2023danse,yang2023low, juarez2022implementation}. Doing so  bypasses the need to estimate the missing aspects of the  \ac{ss} model, learning to track via end-to-end training. In particular, the KalmanNet architecture~\cite{revach2022kalmannet} and its variants~\cite{choi2023split,ni2022rtsnet,wang2024nonlinear,song2024practical} substitute the \ac{kg} computation of the \ac{ekf} with a  \ac{dnn}. The resulting algorithm was shown to enable accurate  tracking in partially known and complex dynamics, while facilitating combination with  pre-processing~\cite{buchnik2023latent} and downstream state post-processing~\cite{LQGNet_ICASSP, putri2023data, milstein2024neural}. However, as such methods learn to estimate the state,  they do not naturally provide the error covariance, for which there is rarely "ground-truth". 

Uncertainty extraction is typically quite challenging in generic deep learning~\cite{russell2021multivariate}. Even for classification \acp{dnn}, that are  designed to output a probability mass function, learned probabilities are often not faithful for uncertainty extraction~\cite{wei2022mitigating}, while in estimation tasks, generic \acp{dnn} do not quantify uncertainty.  For \ac{dnn}-aided \acp{kf}, several specific extensions were suggested to provide error covariance. For instance, the \ac{rkn} proposed in~\cite{becker2019recurrent} was trained to output an error estimate in addition to the state estimate, while~\cite{klein2022uncertainty} showed that when applying KalmanNet of~\cite{revach2022kalmannet} in linear \ac{ss} models with full column rank observation matrix and Gaussian measurement noise, the error covariance can be extracted from internal features of the architecture. These specific studies motivate a unifying study on error covariance extraction in \ac{dnn}-aided \acp{kf}. 

In this work we study discriminative \ac{dnn}-aided tracking with uncertainty quantification for both linear and non-linear partially known \ac{ss} models. We extensively examine the extraction of error covariance from existing architectures. We categorize \acp{dnn} that can learn to track in partially known \ac{ss} model into model-based deep learning designs, where one can relate internal features to the \ac{kg} or prior covariance, and black-box architectures, where only the output has an operational meaning. We show how the desired error covariance can be extracted in each setting: 
for black-box \acp{dnn}, error covariance estimation necessitates providing additional output neurons; for architectures whose internal features can be attributed to the prior covariance and the \ac{kg}, we identify how and in which conditions can its internal features be mapped into error covariance prediction. We present two training methods which encourage such architectures to support error covariance estimation. 
%
%%%%%%%%%%%%%%%%
%%% Analysis %%%
%%%%%%%%%%%%%%%%
%

Our analysis identifies that reliable extraction of the error covariance from existing \ac{dnn}-aided algorithms is  feasible only in some settings, e.g., when tracking fewer state variables than observed ones. Moreover, doing so requires  additional knowledge of the \ac{ss} model that is not required for state tracking. To overcome this, we propose a novel \ac{dnn}-aided tracking algorithm that supports uncertainty quantification termed {\em Bayesian KalmanNet}. Our proposed algorithm draws inspiration from Bayesian deep learning~\cite{kendall2017uncertainties, jospin2020hands, raviv2023modular}, deviating from conventional {\em frequentist} learning to augment an \ac{ekf} with a {\em stochastic \ac{dnn}} to compute the \ac{kg}. Doing so enables tracking the state with the same level of partial \ac{ss} modeling needed  by conventional KalmanNet and its variants, while providing reliable uncertainty measures by sampling multiple \ac{dnn} realizations. % at each time instance. 

%as in \ac{skn}~\cite{choi2023split}. We also discuss how  error covariance is extracted without modifying the system; and for algorithms learning only the \ac{kg}, e.g., \cite{revach2022kalmannet}, we show how it can be used to obtain the prior covariance, extending the findings of~\cite{klein2022uncertainty} to non-linear \ac{ss} models.
%and whether doing so necessitates modifying the training procedure.
%

We compare Bayesian KalmanNet and our extension of existing \ac{dnn}-aided \acp{kf} for uncertainty extraction in  a qualitative and a quantitative manner. For the latter we consider both a synthetic and a physically compliant navigation setting. Our results  show that Bayesian KalmanNet consistently provides reliable tracking of the state along with faithful error covariance prediction under different levels of accuracy in knowledge of the dynamics, and that incorporating  domain knowledge in model-based deep learning designs systematically yields the most reliable estimate of the error covariance.

% \textbf{Paragraph 4: High level contribution of the work}
% From \ac{mb} to end-to-end to end \ac{dnn}s, this work suggest various ways of extracting the uncertainty from a wide range of deep learning incorporated algorithms. By better quantifying an estimated states' uncertainties, the accuracy and robustness of the estimation is improved and has the potential to enhance the performance of diverse systems such as autonomous vehicles, robotics, and other tracking systems.

% \textbf{Paragraph 5: Go over what is done in the paper}
% Our work provides both a review of recent emerging methodologies that integrate deep-learning techniques with Kalman Filtering and suggest different approaches for extracting uncertainties of the various methods. We perform a numerical evaluation and compare the suggested approaches using two different settings and discuss each method's dependencies and capabilities.

% \textbf{Paragraph 6:} Organization:
The rest of this paper is organized as follows: Section~\ref{sec:system_model} briefly reviews \ac{ss} models and the \ac{ekf}. Extensions of existing \ac{dnn}-aided \acp{kf} for uncertainty extraction are presented in Section~\ref{sec:DeepFrequent}. Bayesian KalmanNet is derived in Section~\ref{sec:BKN},  Section~\ref{sec:results} details our numerical study, and Section~\ref{sec:conclusion} provides concluding remarks.

Throughout this paper, we use  boldface lowercase for vectors, e.g., $\gvec{x}$, and boldface uppercase letters, e.g., $\pmb{M}$, for matrices. The $i$th element of some vector $\gvec{x}$ is denoted $[\gvec{x}]_i$. Similarly, $[\pmb{M}]_{i,j}$ is the $(i,j)$th element of a matrix $\pmb{M}$.
We use $\mathbb{N}$ and $\mathbb{R}$ for the sets of  natural and real numbers, respectively. The operations $(\cdot)^\top$,  $\| \cdot \|$,  $\mathbb{E}\{\cdot \}$, and ${\rm Var}\{\cdot\}$  are used for transpose,  $\ell_2$ norm, stochastic expectation, and variance, respectively.

%%%%%%%%%%%%%%%%%%%%%%%%%%%%%%%%%%%%%%%%%%%%%%%%%%%%%%%%%%%%%%%%%%%%%%%%%%%%%%%%
%\vspace{-0.2cm}
\section{System Model and Preliminaries}
\label{sec:system_model}
%\vspace{-0.15cm}
In this section we review the system model and recall some  preliminaries. We first present the \ac{ss} model in Subsection~\ref{ssec:SSModel} and discuss how it is used for model-based and \ac{dnn}-based tracking in Subsections~\ref{ssec:MBtracking}-\ref{ssec:Deeptracking}, respectively. We formulate the problem of uncertainty-aware tracking in Subsection~\ref{ssec:Problem Formulation}, and review basics in Bayesian deep learning in Subsection~\ref{ssec:Bayesian DL}.

\subsection{SS Model}
\label{ssec:SSModel}
We consider a dynamic system described as a non-linear \ac{ss} model in discrete time. This model describes the relationship between  a latent  state vector  $\gvec{x}_t \in \mathbb{R}^{m}$ and  observation vector   $\gvec{y}_t \in \mathbb{R}^{n}$ for each time instance $t \in \mathbb{N}$. In a non-linear \ac{ss} model, this relationship is generally represented as
\begin{subequations}
\label{eq:gen_state}
\begin{align}
    \gvec{x}_t &= \gvec{f}(\gvec{x}_{t-1}) + \pmb{w}_t, 
    \label{eq:gen_state_evo} \\
    \gvec{y}_t &= \gvec{h}(\gvec{x}_{t}) + \pmb{v}_t.  
    \label{eq:gen_state_obs} 
\end{align}
\end{subequations}
In  \eqref{eq:gen_state}, $\gvec{f}:  \mathbb{R}^{m} \to \mathbb{R}^{m}$ is the state evolution function; $\gvec{h}:  \mathbb{R}^{m} \to \mathbb{R}^{n}$ is the observation function; and $\pmb{w}_t, \pmb{v}_t$  are mutually and temporally independent zero-mean noises with covariances $\gvec{Q}$ and $\gvec{R}$, respectively.

While various tasks are associated with \ac{ss} models, including, e.g., smoothing and imputation, we focus on {\em tracking}.
The tracking (filtering) task  refers to estimating the state $\gvec{x}_t$ from current and past observations $\{  \gvec{y}_\tau\}_{\tau\leq t}$. Performance is measured via the \ac{mse} between the estimated state and the true one. Namely, for a sequence of length $T$, the time-averaged \ac{mse} of an estimate $\{\hat{\gvec{x}}_t\}_{t=1}^T$ is
\begin{equation}
\label{eq:MSE}
    {\rm MSE} = \frac{1}{T}\sum_{t=1}^T\mathbb{E}\left\{\left\| \hat{\gvec{x}}_t - {\gvec{x}}_t\right\|^2 \right\}.
\end{equation}

\subsection{Model-Based Tracking}
\label{ssec:MBtracking}
The characterization of dynamic systems via  the \ac{ss} model  \eqref{eq:gen_state} gives rise to various algorithms for tracking the state. 
Specifically, given a full and accurate description of the functions $\gvec{f}(\cdot)$, $\gvec{h}(\cdot)$, and characterization of the distribution of the state and measurement noises as being Gaussian with known covariances, a suitable tracking algorithm is the \ac{ekf}~\cite[Ch. 10]{durbin2012time}. 

The \ac{ekf} is comprised of \textit{prediction} and \textit{measurement update} steps. 
On each time step $t$, it starts by propagating the first-order statistical moments of the state and observations  via 
\begin{equation}
\label{eqn:ekf_pred}
 \hat{\gvec{x}}_{t|t-1} = \gvec{f}(\hat{\gvec{x}}_{t-1|t-1}), \quad  \hat{\gvec{y}}_{t|t-1} = \gvec{h}(\hat{\gvec{x}}_{t|t-1}).    
\end{equation}
 %respectively. 
Letting $\hat{\gvec{F}}_{t}$ and $\hat{\gvec{H}}_{t}$ be the Jacobians (partial derivatives) of $\gvec{f}(\cdot)$ and $\gvec{h}(\cdot)$, evaluated at $\hat{\gvec{x}}_{t-1|t-1}$ and $\hat{\gvec{x}}_{t|t-1}$, respectively, the prior covariances are set to
\begin{subequations}
\label{eq:kf_cov_predBoth}
\begin{align}
  %  \hat{\gvec{x}}_{t|t-1} &= \hat{\gvec{F}}_{t}\hat{\gvec{x}}_{t-1|t-1}, \label{eq:kf_state_evo} \\
    \boldsymbol{\Sigma}_{t|t-1} &= \hat{\gvec{F}}_{t}\boldsymbol{\Sigma}_{t-1|t-1}\hat{\gvec{F}}_{t}^\top + \gvec{Q},\label{eq:kf_cov_evo}\\
% \end{align}
% \end{subequations}
% and of the observations as
% \begin{subequations}
% \begin{align}
%     \hat{\gvec{y}}_{t|t-1} &= \hat{\gvec{H}}_{t}\hat{\gvec{x}}_{t-1|t-1}, \label{eq:kf_obs_evo} \\
    \gvec{S}_{t|t-1} &= \hat{\gvec{H}}_{t}\boldsymbol{\Sigma}_{t|t-1}\hat{\gvec{H}}_{t}^\top + \gvec{R}.\label{eq:kf_obs_cov}
    \vspace{-0.1cm}
\end{align}
\end{subequations}
The observations are fused with the predictions as 
\begin{equation}
\label{eq:EKFupdate}
   \hat{\gvec{x}}_{t|t} = \hat{\gvec{x}}_{t|t-1} + \gvec{K}_{t}( {\gvec{y}}_{t} - \hat{\gvec{y}}_{t|t-1}), 
\end{equation}
where $\gvec{K}_{t}$ is the \ac{kg} given by
\begin{equation}\label{eq:kf_gain_calc}
    \gvec{K}_{t} = \boldsymbol{\Sigma}_{t|t-1}\hat{\gvec{H}}_{t}^\top\gvec{S}^{-1}_{t|t-1}.
\end{equation}
The error covariance is then computed as  
\begin{align}
%    \hat{\gvec{x}}_{t|t} &= \hat{\gvec{x}}_{t|t} + \gvec{K}_{t}\cdot\Delta{\gvec{y}}_{t}; \quad \Delta{\gvec{y}}_{t} := {\gvec{y}}_{t} - \hat{\gvec{y}}_{t|t-1},\label{eq:kf_state_upd} \\
    \boldsymbol{\Sigma}_{t|t} &= (\gvec{I}-  \gvec{K}_{t} \hat{\gvec{H}}_{t}) \boldsymbol{\Sigma}_{t|t-1}.  \label{eq:kf_cov_upd}
\end{align} 
% where $\gvec{K}_{t}$ is the \ac{kg} computed by
% \begin{equation}\label{eq:kf_gain_calc}
%     \gvec{K}_{t} = \hat{\boldsymbol{\Sigma}}_{t|t-1}\hat{\gvec{H}}_{t}^\top\gvec{S}^{-1}_{t|t-1}.
% \end{equation}

The \ac{ekf} copes with the non-linearity of the \ac{ss} model by linearizing the state evolution and measurement functions when propagating the second-order moments in \eqref{eq:kf_cov_predBoth}. Alternative approaches employ different approximations for propagating these moments in non-linear \ac{ss} models, including  the unscented transform (utilized for that purpose by the  \ac{ukf} ~\cite{julier2004unscented}),    as well as  the cubature~\cite{arasaratnam2009cubature} and Gauss-Hermite deterministic or stochastic quadrature rules~\cite{arasaratnam2007discrete,Dunik2015}.
While these model-based  algorithms  may not necessarily be \ac{mse} optimal (as opposed to linear Gaussian \ac{ss} models), they $(i)$ inherently provide an estimate of both the state via \eqref{eq:EKFupdate} and the estimation uncertainty via \eqref{eq:kf_cov_upd}; and $(ii)$ rely on  accurate description of the dynamics as a known \ac{ss} model as in \eqref{eq:gen_state}.

% %%%%%%%%%%%%%%%%%%%%%%%%%%%
% %%% DNN-Based Tracking
% %%%%%%%%%%%%%%%%%%%%%%%%%%%
\subsection{DNN-Aided Tracking}
\label{ssec:Deeptracking} 
\Acp{dnn}, and particularly architectures suitable of time sequences such as \acp{rnn}, can learn to track in dynamics where the \ac{ss} model is complex and unknown \cite{lim2021time}.   
\acp{dnn} used for tracking take as input the current observation $\gvec{y}_t$ at each time $t \in \mathbb{N}$. The  output of such a \ac{dnn}-aided filter with trainable parameters $\dnnParams$ at time $t$, denoted $\hat{\gvec{x}}_t(\gvec{y}_t,\dnnParams) \in \mathbb{R}^m$, is an  estimate of the state $\gvec{x}_t$, as illustrated in Fig.~\ref{fig:Approaches_Scheme}(A).

We focus on supervised settings, where learning is based on   data  comprised of  pairs of state length $T$ state trajectories and  observations, denoted 
\begin{equation}
\label{eq:dataset}
   \mySet{D} = \big\{\{\gvec{x}_t^{(i)},\gvec{y}_t^{(i)}\}_{t=1}^T\big\}_{i=1}^{|\mySet{D}|}, 
\end{equation}
 where the superscript $(i)$ denotes $i$th data trajectory. In training, the \ac{dnn} parameters  $\dnnParams$ are tuned such that the state estimate best matches the data. Since performance is measured via the \ac{mse} \eqref{eq:MSE}, a suitable loss function is  the empirical $\ell_2$ loss 
\begin{equation}
    \label{eqn:Loss}
    \mySet{L}_{\mySet{D}}^{\ell_2}(\dnnParams) = \frac{1}{|\mySet{D}| T}\sum_{i=1}^{|\mySet{D}|} \sum_{t=1}^T \big\|\gvec{x}^{(i)}_{t}-\hat{\gvec{x}}_{t}(\gvec{y}^{(i)}_t,\dnnParams)\big\|^2. 
\end{equation}
While \acp{dnn} can be trained using \eqref{eqn:Loss} to track $\gvec{x}_t$ without requiring characterization of the \ac{ss} model in \eqref{eq:gen_state}, they do not  provide an estimate of the error covariance, e.g.,  as the \ac{ekf} computes $\boldsymbol{\Sigma}_{t|t}$ in \eqref{eq:kf_cov_upd}. Existing approaches for explainability in \acp{dnn}, typically involve additional data and conformal prediction tools for boosting verification within  confidence regions~\cite{lindemann2024formal}, rather than providing instantaneous error covariance.
%
%As mentioned in \cite{lindemann2024formal}, incorprating conformal prediction could also allow uncertainty quantification in \ac{dnn}-aided \acp{kf} by generating a statistically valid prediction region that encapsulates the true state with high probability.

% %%%%%%%%%%%%%%%%%%%%%%%%%%%
% %%% Problem Formulation
% %%%%%%%%%%%%%%%%%%%%%%%%%%%
\subsection{Problem Formulation}
\label{ssec:Problem Formulation} 
We consider the problem of {\em uncertainty-aware tracking}, namely, the design of an algorithm that maps the observed $\gvec{y}_t$ into estimates of both the state $\hat{\gvec{x}}_t$ and its {\em estimation uncertainty} $\hat{\boldsymbol{\Sigma}}_t$. The former is evaluated via the \ac{mse} in \eqref{eq:MSE}, while the latter is measured via the accuracy of the estimated variances, i.e.,
\begin{equation}
    \frac{1}{T}\sum_{t=1}^T\sum_{j=1}^m\mathbb{E}\left\{\left| \left[\hat{\boldsymbol{\Sigma}}_t\right]_{j,j} - {\rm Var}\left\{\left[{\gvec{x}}_t\right]_j \right\}\right| \right\}.
\end{equation}

Model-based tracking algorithms estimate both the state and the error covariance when the dynamics are characterized by a known \ac{ss} model. However, we focus on setting where the \ac{ss} is only partially known, and one must cope with the following requirements:
\begin{enumerate}[label={R\arabic*}]
\item \label{itm:FuncAccuracy} The state evolution and observation models,  $\gvec{f}(\cdot)$ and $\gvec{h}(\cdot)$, are {\em approximations of the true dynamics}, that are possibly mismatched. 
\item \label{itm:Stochasticity} The distribution of the noises $\myVec{v}_t$ and $\myVec{w}_t$, which capture the inherent stochasticity in the dynamics, is unknown, and can be complex and non-Gaussian.  
\item \label{itm:Error} As  error covariance is a statistical quantity, it is unlikely to be  provided in a dataset $\mySet{D}$. Accordingly, we assume access to data that is labeled as in \eqref{eq:dataset}, namely, it is comprised of trajectories of observations and states but does not include the error covariance.
\end{enumerate}

These requirements motivate exploring data-driven approaches for tackling tasks associated with \ac{ss} models, allowing to both recover the state and the error covariance.

% %%%%%%%%%%%%%%%%%%%%%%%%%%%
% %%% Bayesian Deep Learning
% %%%%%%%%%%%%%%%%%%%%%%%%%%%
\subsection{Bayesian Deep Learning}
\label{ssec:Bayesian DL} 

% https://jorisbaan.nl/2021/03/02/introduction-to-bayesian-deep-learning.html
\textcolor{NewColor}{While the formulation of the \ac{ss} model in Subsection~\ref{ssec:SSModel} and the model-based tracking in Subsection~\ref{ssec:MBtracking} focuses on a Bayesian setting, from a deep learning perspective, the term Bayesian deep learning refers to a framework in which a stochastic modeling is employed on the \ac{dnn} itself. Accordingly, we henceforth use the term {\em frequentist} within the context of \acp{dnn} to refer to the standard deep learning approach in which the weights of a \ac{dnn} are deterministic weights, whereas a {\em Bayesian} \ac{dnn} models its weights as random variables governed by a learned posterior distribution~\cite{kendall2017uncertainties}.}

Specifically, Bayesian deep learning offers a framework to quantify uncertainty in \ac{dnn} predictions. It treats the \ac{dnn} parameters as random variables, and learns their posterior distribution given the data. This probabilistic approach provides a measure of uncertainty by sampling multiple realizations of the (random) \ac{dnn} during inference~\cite{wilson2020bayesian}. % in state estimates and can be particularly useful in tracking applications where error covariance is important.

In Bayesian deep learning, instead of learning a single set of parameters $\dnnParams$, we aim to learn the posterior distribution $p(\dnnParams|\mySet{D})$ over the parameters given the data $\mySet{D}$. In the considered context of estimating a state $\gvec{x}_t$ from observations $\gvec{y}_t$, the state estimate $\hat{\gvec{x}}_t$ and the associated uncertainty can then be obtained by marginalizing over this posterior distribution, i.e.,
% \vspace{-0.1cm}
\begin{equation}
\label{eq:Posterior Distribution}
p(\hat{\gvec{x}}_t|\gvec{y}_t, \mySet{D}) = \int p(\hat{\gvec{x}}_t|\gvec{y}_t, \dnnParams) p(\dnnParams|\mySet{D})d\dnnParams.
% \vspace{-0.1cm}
\end{equation} 
% and similarly, the statistical moments of the estimated state are obtained as
% \begin{equation}
% \label{eq:Posterior mean}
% \mathbb{E}\{\hat{\gvec{x}}_t|\gvec{y}_t, \mySet{D}\} = \int \mathbb{E}\{\hat{\gvec{x}}_t|\gvec{y}_t\hat{\gvec{x}}_t|\gvec{y}_t, \dnnParams\} p(\dnnParams|\mySet{D})d\dnnParams,
% % \vspace{-0.1cm}
% \end{equation} 

To learn a stochastic \ac{dnn}, one typically seeks \textcolor{NewColor2}{to approximate a posterior} distribution on the weights, i.e., $p(\dnnParams|\mySet{D})$, that is dictated by some parameters $\bnnParams$. This can result in either an explicit formulation of the distribution, as in variational inference~\cite{fortuin2022priors}, or via alternative induced trainable stochasticity, as in Monte Carlo dropout~\cite{gal2017concretedropout}. Once dictated, the parameters of the \ac{dnn} distribution, $\bnnParams$, are trained using the \ac{elbo} objective, which regularizes the expected loss with a term minimizing the \ac{kl} divergence between the induced distribution $q(\dnnParams|\bnnParams)$ and some pre-determined prior $p(\dnnParams)$, i.e., 
\begin{equation}
\label{eq:ELBO}
\text{ELBO}_{\mySet{D}}(\bnnParams) = \mathbb{E}_{q(\dnnParams|\bnnParams)} \left[ \log p(\mySet{D}|\dnnParams) \right] - \text{KL}(q(\dnnParams|\bnnParams) | p(\dnnParams)).
% \vspace{-0.1cm}
\end{equation}

Once trained, the stochastic \ac{dnn} is applied via ensembling method. Particularly, during inference $J$ realizations, denoted $\{\dnnParams_j\}_{j=1}^J$, are drawn in an i.i.d. manner from $q(\dnnParams|\bnnParams)$. Then, inference is carried out by applying these $J$ \acp{dnn} to the input, and combining their results. For instance, in our context of state tracking, the resulting estimate becomes 
\begin{equation}
\label{eqn:BNNInference}
\hat{\gvec{x}}_t(\myVec{y}_t, \bnnParams) = \frac{1}{J}\sum_{j=1}^{J} \hat{\gvec{x}}_{t}^{(j)}(\gvec{y}_t,\dnnParams_j), \quad \{\dnnParams_j \}\stackrel{\rm i.i.d.}{\sim} q(\dnnParams|\bnnParams). 
\end{equation}

\section{Error Covariance Extraction in Frequentist DNN-Aided Tracking}
\label{sec:DeepFrequent}
%\vspace{-0.15cm}
The first part of our study focuses on how one can extract  error covariance from existing \ac{dnn}-based tracking algorithms, based on conventional frequentist training (i.e., learning of deterministic \acp{dnn}). 
We divide existing architectures by their interpretable features that can be related to error covariance in Subsection~\ref{ssec:Architecture}. Then, we propose dedicated training objectives to encourage these architectures to allow extraction of faithful uncertainty measures in Subsection~\ref{ssec:Training}, respectively, and provide a discussion in Subsection~\ref{ssec:discussion}.

%%%%%%%%%%%%%%%%%%%%%%%%%%%
%%% Error Covariance Indicative Features
%%%%%%%%%%%%%%%%%%%%%%%%%%%
% \vspace{-0.1cm}
\subsection{Error Covariance Indicative Features}
\label{ssec:Architecture}
% \vspace{-0.1cm}
Different \ac{dnn}-aided algorithms provide different features that can be used to extract the error covariance. We divide \ac{dnn}-aided systems accordingly, considering black-box methods, where only the outputs are observed, as well as model-based deep learning algorithms following the \ac{ekf} operation, where one can possibly relate internal features to the prior covariance and  \ac{kg}.
%in black-box architectures (that are invariant of the \ac{ss} model) and model-based deep learning techniques (that incorporate partial domain knowledge). 

\subsubsection{Output Features} 
In conventional black-box \acp{dnn}, one can only assign an interpretable meaning to the input and output of the system. In such cases, in order to estimate the error covariance, additional outputs are required (see Fig.~\ref{fig:Approaches_Scheme}(B)); these outputs should represent the estimated covariance denoted $\hat{\boldsymbol{\Sigma}}_t$. This can be achieved by having the \ac{dnn} output (in addition to $\hat{\gvec{x}}_t(\gvec{y}_t,\dnnParams)$), a vector $\hat{\boldsymbol{\Sigma}}_t(\gvec{y}_t,\dnnParams)\in \mathbb{R}_+^m$, as done in  the \ac{rkn} method \cite{becker2019recurrent},  setting the estimated error covariance to be
\begin{equation}
\label{eqn:ErrorCovRKN}
    \hat{\boldsymbol{\Sigma}}_t(\gvec{y}_t,\dnnParams) = {\rm diag}(\hat{\boldsymbol{\Sigma}}_t(\gvec{y}_t,\dnnParams)).
\end{equation}

\subsubsection{KG and Prior Covariance Features} 
Several hybrid model-based deep learning learn to track in partially known \ac{ss} models by augmenting the \ac{ekf} with  \acp{dnn}. Some algorithms, such as \ac{skn} \cite{choi2023split}, learn to compute the \ac{kg}, denoted $\hat{\gvec{K}}_t(\gvec{y}_t,\dnnParams)$, by treating an internal feature as an estimate of the prior covariance, denoted $\hat{\boldsymbol{\Sigma}}_{t|t-1}(\gvec{y}_t,\dnnParams)$. Following the  \ac{ekf} \eqref{eq:kf_cov_predBoth}--\eqref{eq:kf_cov_upd}, these features can be combined to estimate the error covariance as illustrated in  Fig.~\ref{fig:Approaches_Scheme}(C) via 
\begin{equation}
\label{eqn:ErrorCovSKN}
    \hat{\boldsymbol{\Sigma}}_t(\gvec{y}_t,\dnnParams) = (\gvec{I}-  \hat{\gvec{K}}_t(\gvec{y}_t,\dnnParams) \hat{\gvec{H}}_{t}) \hat{\boldsymbol{\Sigma}}_{t|t-1}(\gvec{y}_t,\dnnParams).
\end{equation}

The formulation in \eqref{eqn:ErrorCovSKN} is based on the update step of the \ac{ekf} in \eqref{eq:kf_cov_upd}, which implicitly assumes that the \ac{kg} is the \ac{mse} optimal linear estimator. As there is no guarantee that this indeed holds for a learned \ac{dnn}-aided \ac{kg}, one can still estimate the error covariance via the Joseph form
%
% \begin{align}
% &\hat{\boldsymbol{\Sigma}}_t(\gvec{y}_t,\dnnParams) =  \hat{\gvec{K}}_t(\gvec{y}_t,\dnnParams)\cdot
% \gvec{R}
% \cdot
% \hat{\gvec{K}}_t^\top(\gvec{y}_t,\dnnParams) + \\ \notag
% %
% &(\gvec{I}-  \hat{\gvec{K}}_t(\gvec{y}_t,\dnnParams) \hat{\gvec{H}}_{t}) 
% %
% \cdot \hat{\boldsymbol{\Sigma}}_{t|t-1}(\gvec{y}_t,\dnnParams) 
% \cdot
% (\gvec{I}-  \hat{\gvec{K}}_t(\gvec{y}_t,\dnnParams) \hat{\gvec{H}}_{t})^\top .
% \label{eqn:ErrorCovJoseph}
% \end{align}
\begin{align}
    \hat{\boldsymbol{\Sigma}}_t(\gvec{y}_t,\dnnParams) &=  \hat{\gvec{K}}_t(\gvec{y}_t,\dnnParams)\gvec{R}\hat{\gvec{K}}_t^\top(\gvec{y}_t,\dnnParams) + (\gvec{I}-  \hat{\gvec{K}}_t(\gvec{y}_t,\dnnParams) \hat{\gvec{H}}_{t})  \notag \\
    &\times \hat{\boldsymbol{\Sigma}}_{t|t-1}(\gvec{y}_t,\dnnParams) (\gvec{I}-  \hat{\gvec{K}}_t(\gvec{y}_t,\dnnParams) \hat{\gvec{H}}_{t})^\top .
\label{eqn:ErrorCovJoseph}
\end{align}

\subsubsection{KG Features} 
The formulations in \eqref{eqn:ErrorCovSKN}-\eqref{eqn:ErrorCovJoseph} rely on the ability to extract both the prior covariance and the \ac{kg} as internal features. \ac{dnn}-aided systems such as KalmanNet~\cite{revach2022kalmannet} learn to compute the \ac{kg}  from data without recovering the prior covariance. There, if  $\Tilde{\gvec{H}}_{t} \triangleq \big(\hat{\gvec{H}}_{t}^\top\hat{\gvec{H}}_{t}\big)^{-1}$ exists,  then, by  combining \eqref{eq:kf_obs_cov}-\eqref{eq:kf_gain_calc},   the prior covariance can be recovered via 
\begin{equation}
\label{eqn:PriorErrorCov}
      \! \hat{\boldsymbol{\Sigma}}_{t|t\!-\!1}(\gvec{y}_t,\dnnParams) \!=\! (\gvec{I} \!- \!\hat{\gvec{K}}_t(\gvec{y}_t,\dnnParams) \hat{\gvec{H}}_t)^{-1}\hat{\gvec{K}}_t(\gvec{y}_t,\dnnParams) \gvec{R}\hat{\gvec{H}}_t\Tilde{\gvec{H}}_{t}.
\end{equation}

\begin{figure}
\includegraphics[width=\linewidth]{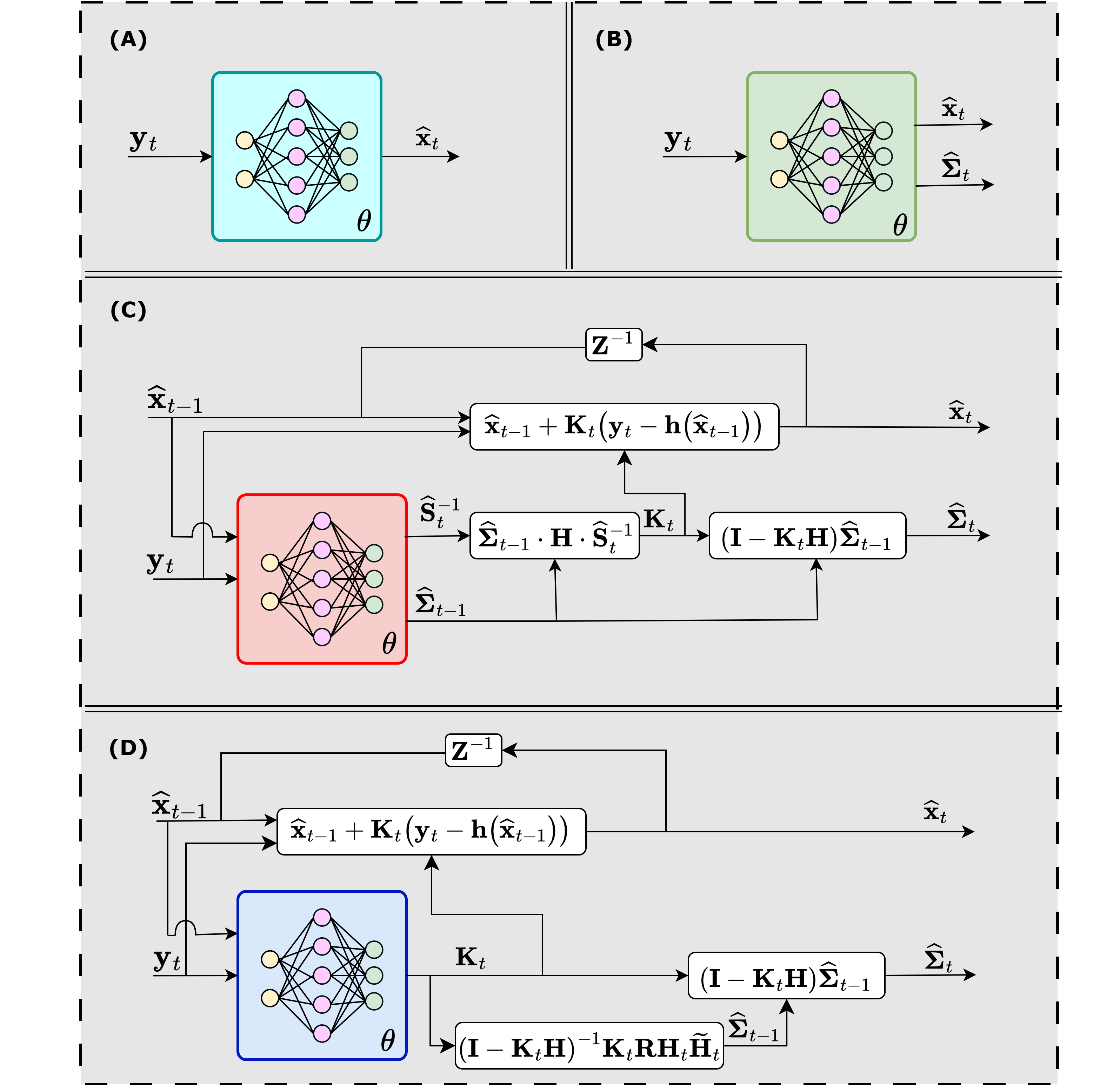}
% \includegraphics[width=\linewidth]{figs/Approaches_Scheme.jpg}
%\vspace{-0.7cm}
\caption{Considered tracking approaches: $(A)$ standard black-box \ac{dnn}; $(B)$ \ac{dnn} with error covariance output (e.g., \cite{becker2019recurrent}); $(C)$ \ac{dnn}-aided \ac{kf}, estimated posterior covariance and \ac{kg} (e.g., \cite{choi2023split}); and $(D)$ \ac{dnn}-aided \ac{kf}, learned \ac{kg} (e.g., \cite{revach2022kalmannet}).}
%\vspace{-0.4cm}
\label{fig:Approaches_Scheme}
\end{figure}

Using the recovered prior covariance in \eqref{eqn:PriorErrorCov}, the desired error covariance is obtained by either \eqref{eqn:ErrorCovSKN} or \eqref{eqn:ErrorCovJoseph}. This approach is illustrated in  Fig.~\ref{fig:Approaches_Scheme}(D).

\begin{remark}
\label{rmk:Partial}
Converting the \ac{kg} into the prior covariance in \eqref{eqn:PriorErrorCov} requires $\Tilde{\gvec{H}}_{t}$ to exist, i.e., $\hat{\gvec{H}}_{t}$ has full column rank. Yet, even when  some of the columns of $\hat{\gvec{H}}_{t}$ are zero or $n<m$, error covariance matrix elements with respect to the same state variables can still be recovered by considering the sub-matrices of $\hat{\gvec{H}}_{t}$, $ \hat{\boldsymbol{\Sigma}}_{t|t-1}$, and $ \hat{\boldsymbol{\Sigma}}_{t}$ corresponding to these variables. For instance,  in navigation scenarios with noisy position measurements, one can still extract the position error covariance while also tracking velocity.%, as  shown in our experimental study in Section~\ref{sec:results}.
\end{remark}

%%%%%%%%%%%%%%%%%%%%%%%%%%%
%%% Training to Predict Error Covariance
%%%%%%%%%%%%%%%%%%%%%%%%%%%
%\vspace{-0.4cm}
\subsection{Training to Predict the Error Covariance}
\label{ssec:Training}
% \vspace{-0.1cm}
The indicative features detailed in Subsection~\ref{ssec:Architecture} allow \ac{dnn}-aided systems to provide $ \hat{\boldsymbol{\Sigma}}_t(\gvec{y}_t,\dnnParams)$,  viewed as an estimate of the error covariance. 
However, training via \eqref{eqn:Loss}, which solely observes the state estimate, does not explicitly encourage $ \hat{\boldsymbol{\Sigma}}_t(\gvec{y}_t,\dnnParams)$ to accurately match the error covariance. Accordingly, we consider two alternative loss measures that overcome the absence of a ground truth covariance, one that is based on evaluating the empirically averaged error variance, and one that utilizes a Gaussian prior of the estimated state.
%
%%%%%%%%%%%%%%
%%% Figure %%%
%%%%%%%%%%%%%%
%
%
%%%%%%%%%%%%%%%%%%%%%%%%%%%%%%%%
%%% Empirical Error Variance %%%
%%%%%%%%%%%%%%%%%%%%%%%%%%%%%%%%
%
\subsubsection{Empirical Error Variance} 
A straight-forward approach to account for the error covariance in the loss  is to introduce an additional term that compares $ \hat{\boldsymbol{\Sigma}}_t(\gvec{y}_t,\dnnParams)$ to the instantaneous empirical estimate. Specifically, by defining the error vector 
\begin{equation}
\label{eqn:errorVec}
    \pmb{e}_t(\gvec{y}_t, \gvec{x}_t, \dnnParams) \triangleq \gvec{x}_{t}-\hat{\gvec{x}}_{t}(\gvec{y}_t,\dnnParams),
\end{equation}
%and  $\hat{\boldsymbol{\Sigma}}_t(\gvec{y}_t,\dnnParams)$ as the diagonal vector of $\hat{\boldsymbol{\Sigma}}_t(\gvec{y}_t^{(i)},\dnnParams)$, % as 
% \begin{equation*}
%     \hat{\pmb{C}}^{\mySet{D}}_t(\dnnParams) = \frac{1}{|\mySet{D}|}\sum_{i=1}^{|\mySet{D}|}\left(\gvec{x}^{(i)}_{t}-\hat{\gvec{x}}_{t}(\gvec{y}^{(i)}_t,\dnnParams) \right)\left(\gvec{x}^{(i)}_{t}-\hat{\gvec{x}}_{t}(\gvec{y}^{(i)}_t,\dnnParams) \right)^\top,
% \end{equation*}
the  loss includes an additive term that compares the predicted error with its empirical value obtained from \eqref{eqn:errorVec}. By defining the second moment loss over the whole dataset as:
\begin{equation}\label{eqn:M2Loss}
\mySet{L}_{\mySet{D}}^{\rm M2}(\dnnParams) =  \sum_{i=1}^{|\mySet{D}|}\sum_{t=1}^T \sum_{j=1}^{m} \left|[\pmb{e}_t(\gvec{y}^{(i)}_t,\gvec{x}^{(i)}_t,\dnnParams)]_j^2 - [\hat{\boldsymbol{\Sigma}}_t(\gvec{y}_t^{(i)},\dnnParams)]_{j,j}\right|.
\end{equation}
The resulting empirical loss used for training is 
\begin{equation}\label{eqn:EmpAvgLoss}
\mySet{L}_{\mySet{D}}^{\rm Emp}(\dnnParams) =  
(1-\beta)\cdot\mySet{L}_{\mySet{D}}^{\ell_2}(\dnnParams) +
\beta\cdot\mySet{L}_{\mySet{D}}^{\rm M2}(\dnnParams).
\end{equation}
\textcolor{NewColor}{The hyperparameter $\beta\in [0,1]$ in~\eqref{eqn:EmpAvgLoss} is used to balance the trade-off between minimizing the tracking error and encouraging accurate prediction of the error covariance. In our numerical implementation detailed in Section~\ref{sec:results}, $\beta$ is initialized with a small value at the beginning of training, so that the network initially focuses on learning to accurately track the state. As training progresses, $\beta$ is gradually increased, thereby placing more emphasis on the accuracy of the predicted uncertainty.}
  Specifically, when setting $\beta=0$, then training with \eqref{eqn:EmpAvgLoss} coincides with training with the standard $\ell_2$ loss. 
%\smallskip

\subsubsection{Gaussian Prior} 
An alternative loss that accounts for error covariance without  explicitly comparing predicted and empirical values  leverages Gaussian priors. Particularly, one can train to both track and provide error covariance by imposing a Gaussian prior on the tracked state, as done in~\cite{becker2019recurrent}. Under this assumption, a suitable loss is the log-likelihood  of a Gaussian distribution with mean $\hat{\gvec{x}}_{t}(\gvec{y}^{(i)}_t,\dnnParams)$ and covariance  $\hat{\boldsymbol{\Sigma}}_t(\gvec{y}_t^{(i)},\dnnParams)$, i.e., % The resulting training loss is
\begin{align}
\mySet{L}_{\mySet{D}}^{\rm GP}(\dnnParams) =  \frac{1}{|\mySet{D}| T}\sum_{i=1}^{|\mySet{D}|} \sum_{t=1}^T &\pmb{e}_t^\top(\gvec{y}^{(i)}_t,\gvec{x}^{(i)}_t,\dnnParams)\hat{\boldsymbol{\Sigma}}_t^{-1}(\gvec{y}_t^{(i)},\dnnParams) \notag \\ &\times \pmb{e}_t(\gvec{y}^{(i)}_t, \gvec{x}^{(i)}_t, \dnnParams) \notag \\
& +\log  \det \left(\hat{\boldsymbol{\Sigma}}_t(\gvec{y}_t^{(i)},\dnnParams)\right).
\label{eqn:GNLL_Loss}
\end{align}
We note that when the error covariance  $\hat{\boldsymbol{\Sigma}}_t$ is not estimated, but is instead assumed to be a scaled identity, then training with the loss in \eqref{eqn:GNLL_Loss} is equivalent to training with the standard $\ell_2$ loss. 

\subsection{Discussion}
\label{ssec:discussion}
% \vspace{-0.1cm}
%\textbf{The discussion is very important. It should provide a qualtitative comparison between the approaches in how each appraoch extract uncertainty and what domain knowledge is needed to do so. You can also add a table to that aim (see, e.g., \cite{shlezinger2022collaborative})}
We identified three different approaches to extract error covariance in discriminative \ac{dnn}-aided tracking in non-linear \ac{ss} models. These approaches vary based on the architecture employed and the level of interpretability it provides. It is emphasized though that for all considered approaches, one must still alter the learning procedure for the extracted features to be considered as uncertainty measures. To encourage learning to accurately predict the error, we propose two  training objectives. This examination of error covariance extraction in \ac{dnn}-aided \acp{kf} combines new approaches, while encompassing techniques used for specific schemes (e.g., \eqref{eqn:GNLL_Loss} used by \ac{rkn} \cite{becker2019recurrent}), and extending previous findings (e.g., \eqref{eqn:PriorErrorCov} extends \cite{klein2022uncertainty} to non-linear   models).

The main distinction between the  approaches lies in their  domain knowledge, i.e., which parts of the \ac{ss} model are needed to track and provide error covariance, as summarized in Table~\ref{tbl:QualtComp}. Specifically, black-box \acp{dnn} do not require any model knowledge. Nonetheless, as empirically demonstrated in Section~\ref{sec:results}, using black-box architectures with additional outputs regarded as the error covariance often leads to inaccurate tracking and unreliable error prediction, even when trained with a covariance-oriented loss as the ones proposed in Subsection~\ref{ssec:Training}. 
Model-based deep architectures that learn the \ac{kg} require only approximate knowledge of $\gvec{f}(\cdot)$ and $\gvec{h}(\cdot)$ to track~\cite{choi2023split,revach2022kalmannet}. However, to extract covariance, $\gvec{h}(\cdot)$ should be accurately known (as, e.g., \eqref{eqn:ErrorCovSKN} relies on $\hat{\gvec{H}}_t$), where $\gvec{R}$ is also needed when the prior covariance is not tracked, as in KalmanNet. Accordingly, requirements \ref{itm:FuncAccuracy} and \ref{itm:Stochasticity} are not fully met by these methods.  Moreover, the latter also requires the linearized observation model (i.e., $\tilde{\gvec{H}}_t$) to have full column rank. When these requirements are met, it is empirically shown in Section~\ref{sec:results} that, when training   via the losses proposed in Subsection~\ref{ssec:Training}, one can both accurately track the latent state and reliably predict its covariance from the internal features of the \ac{dnn}-aided algorithm. 
\section{Bayesian KalmanNet}
\label{sec:BKN}
% \vspace{-0.1cm}
As detailed in the previous section, and empirically demonstrated in Section~\ref{sec:results}, model-based deep learning architectures based on the KalmanNet algorithm can simultaneously provide accurate state tracking  alongside reliable error prediction. Nonetheless, achieving the latter requires deviating from how these \ac{dnn}-aided tracking algorithms are designed (e.g., in \cite{revach2022kalmannet} and \cite{choi2023split}) in the following aspects:
\begin{enumerate}[label={A\arabic*}]
    \item \label{itm:domainknowledge} Additional domain knowledge  that is not required merely for state tracking  is needed for covariance extraction (e.g., full characterization of the observations model~\eqref{eq:gen_state_obs}), and the family of dynamic systems where one can extract uncertainty is limited (e.g., $\tilde{\gvec{H}}_t$ has to be non-singular). 
    \item \label{itm:training} The training objective should be altered to encourage reliable uncertainty extraction. 
\end{enumerate}

In this section we propose a \ac{dnn}-aided learning algorithm that preserves the ability of model-based deep learning algorithms such as KalmanNet to accurately track, in partially known \ac{ss} models, while providing uncertainty measures without being subject to the limitations in  \ref{itm:domainknowledge}, thus fully meeting requirements \ref{itm:FuncAccuracy}-\ref{itm:Error}. We build upon the observation in \ref{itm:training}, namely, that error covariance extraction already necessitates altering the learning objective. Instead of just altering the learning objective, we modify the learning {\em paradigm}, deviating from conventional frequentist learning into Bayesian learning of stochastic \acp{dnn}. Our proposed algorithm, coined {\em Bayesian KalmanNet}, is detailed in Subsection~\ref{ssec: BKN Architectecture}, with its training procedure formulated in Subsection~\ref{ssec: Training Bayesian KalmanNet}, while Subsection~\ref{ssec: reduced domain knowledge} provides a discussion.

% In this section, we introduce \ac{bkn}, a novel approach that extends KalmanNet by incorporating Bayesian deep learning techniques to estimate uncertainty. This method leverages Monte Carlo dropout for uncertainty quantification, enabling robust state estimation with reduced domain knowledge.

%%%%%%%%%%%%%%%%%%%%%%%%%%%
%%% Bayesian KalmanNet Tracking Algorithm
%%%%%%%%%%%%%%%%%%%%%%%%%%%
\subsection{Bayesian KalmanNet Tracking Algorithm}
\label{ssec: BKN Architectecture}
As the name suggests, Bayesian KalmanNet is based on the KalmanNet algorithm of \cite{revach2022kalmannet}, while employing Bayesian \acp{dnn}. Accordingly, the algorithm uses a stochastic \ac{dnn} to compute the \ac{kg} of the \ac{ekf}. The usage of a Bayesian \acp{dnn} results in Monte Carlo sampling of $J$ different KalmanNet realizations during inference, that are used to recover both the state and its predicted error via ensemble averaging.

\smallskip
{\bf Architecture:} 
We employ a Bayesian \ac{dnn} with parameters $\dnnParams$ whose distribution is parameterized by $\bnnParams$. Each \ac{dnn}  realization  maps the indicative features $\Delta \gvec{x}_{t-1} \triangleq \hat{\gvec{x}}_{t-1}- \hat{\gvec{x}}_{t-2}$, $\Delta \gvec{y}_t \triangleq \gvec{y}_t - \gvec{y}_{t|t-1}$ and $\Delta \tilde{\gvec{y}}_t \triangleq \gvec{y}_t - \gvec{y}_{t-1}$
 into the  \ac{kg}  $\hat{\gvec{K}}_t(\gvec{y}_t,\dnnParams)$.  

 In our numerical study, we employ {\em Architecture 1} of \cite{revach2022kalmannet}, comprised of \ac{gru} cells with input and output \ac{fc} layers. Stochasticity is achieved by integrating dropout into the \ac{fc} layers~\cite{gal2017concretedropout}, with $\pmb{p}$ denoting the Bernoulli distribution parameter assigned with each neuron of these layers. The  trainable parameters dictating the \ac{dnn} distribution, denoted $\dnnParams$, thus encompass a setting of the \ac{dnn} parameters, denoted $\dnnParams_0$, as well as the dropout neuron-wise probability $\pmb{p}$, i.e., $\bnnParams=[\dnnParams_0, \pmb{p}]$. In each realization, the \ac{dnn} is applied with weights $\dnnParams_0$ with the \ac{fc} layers employing a dropout mask randomized with per-element distribution $\pmb{p}$.

\smallskip
{\bf Uncertainty-Aware Tracking:} 
The proposed augmentation of the \ac{ekf} with Bayesian \acp{dnn} allows to track as in KalmanNet, while leveraging ensembling to predict the estimation error. 
When tracking is launched, we sample $J$ i.i.d. realizations of the \ac{kg} computations \ac{dnn} from $q(\dnnParams|\bnnParams)$, denoted $\{\dnnParams_j\}_{j=1}^J$, with $J$ being a hyperparameter. 
\textcolor{NewColor}{The type of distribution represented by $q(\dnnParams|\bnnParams)$ is a design hyperparameter, as different families of distributions are considered in the Bayesian deep learning literature~\cite{fortuin2022priors}. In our numerical study reported in Section~\ref{sec:results} we employ Monte Carlo dropout~\cite{gal2017concretedropout}, where each \ac{dnn} realization is obtained from the initial fixed weight $\dnnParams_j$ and a realization of the dropout masks sampled with multivariate Bernoulli distribution $\pmb{p}$. However, the formulation of Bayesian KalmanNet is not restricted to Monte Carlo dropout, and can be combined with alternative distributions, e.g., Gaussian weights whose parameters are the first- and second-order moments.}  %Denote, that $J$ is a network's hyperparameter that should be chosen for ad desired task.

On each time instance $t$, the first-order moments of the state and the observation are predicted based on the (possibly approximated) knowledge of $\gvec{f}(\cdot)$ and $\gvec{h}(\cdot)$ via \eqref{eqn:ekf_pred}. These estimated moments are used to form the input features to each of the $J$ \ac{dnn} realizations, obtaining $J$ \acp{kg}, which are in turn used to produce $J$ separate state estimates, denoted $\{\hat{\gvec{x}}_t^j\}_{j=1}^J$. The mean and variance of these estimates provide the state estimate and its associated uncertainty, respectively, via
%
%%%%%%%%%%%%%%%
%%% Moments %%%
%%%%%%%%%%%%%%%
%
\begin{subequations}\label{eqn:BKN}
\begin{align}
\label{eqn:BKNMean}
\hat{\gvec{x}}_t &= \frac{1}{J}\sum_{j=1}^{J}\hat{\gvec{x}}_{t}^{(j)}(\gvec{y}_t,\dnnParams_j),\\
% \end{align}
% and
% \begin{align}
\label{eqn:BKNVar}
\hat{\boldsymbol{\Sigma}}_{t} &= \frac{1}{J} \sum_{j=1}^{J} (\hat{\gvec{x}}_t^{(j)} - \hat{\gvec{x}}_t)(\hat{\gvec{x}}_t^{(j)} - \hat{\gvec{x}}_t)^\top.
%\label{eqn:BKNVar}
\end{align}
\end{subequations}

\begin{figure}
\includegraphics[width=\linewidth]{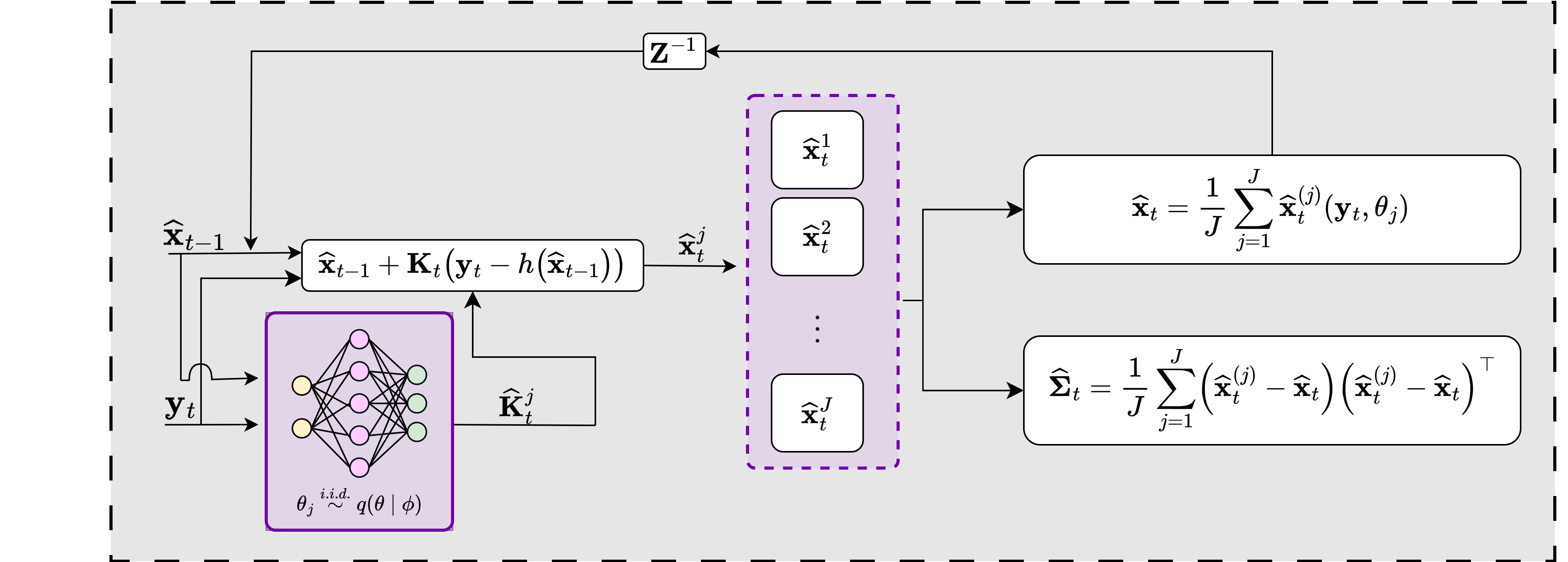}
%\caption{Tracking with Bayesian KalmanNet - Illustration.}
\caption{Bayesian KalmanNet - Architecture}
\label{fig:BKN_Architecture}
\end{figure}

\textcolor{NewColor}{From a Bayesian deep learning perspective, the matrix $\hat{\boldsymbol{\Sigma}}_{t}$ in \eqref{eqn:BKNVar} represents \textcolor{NewColor2}{an empirical estimate of the} \emph{predictive uncertainty}, namely, the second-order moment of the posterior predictive distribution over the state $\mathbf{x}_t$ given the observations and training data~\cite{gawlikowski2023survey}. Specifically, 
each estimate $\hat{\mathbf{x}}_t^{(j)}$ in \eqref{eqn:BKNMean} corresponds to a sample from a different model instantiation, i.e., from the distribution $q(\dnnParams|\bnnParams)$ over the neural network weights. This ensemble of predictions approximates the posterior predictive distribution, and the sample covariance $\hat{\boldsymbol{\Sigma}}_t$ estimates the total predictive uncertainty. While there is no general guarantee that individual estimators $\hat{\mathbf{x}}_t^{(j)}$ are realizations of an unbiased estimator, the usage of its sample covariance for quantifying uncertainty is a standard and widely accepted approach in Bayesian deep learning. Accordingly, our proposed Bayesian KalmanNet uses  $\hat{\boldsymbol{\Sigma}}_{t}$ in \eqref{eqn:BKNVar} as an estimate of the error covariance.}
The resulting operation is summarized as Algorithm~\ref{alg:inference} (in which, for brevity, we write the output of the $j$th \ac{kg} \ac{dnn} realization at time $t$ as $\hat{\gvec{K}}^j_t$), and illustrated in Fig.~\ref{fig:BKN_Architecture}. 
%
%%%%%%%%%%%%%%%%%%%%%%%%%%%%%%%%%%%%%%%%%
%%% Bayesian KalmanNet - Architecture %%%
%%%%%%%%%%%%%%%%%%%%%%%%%%%%%%%%%%%%%%%%%
%
%
%%%%%%%%%%%%%%%%%
%%% Algorithm %%%
%%%%%%%%%%%%%%%%%
%
\begin{algorithm}
\caption{Bayesian KalmanNet}
\label{alg:inference} 
\SetKwInOut{Initialization}{Init}
\Initialization{\ac{dnn} distribution $\bnnParams$;  ensemble size $J$}
\SetKwInOut{Input}{Input} 
\Input{Initial state $\gvec{x}_0$}  
{
Set $\hat{\gvec{x}}_0^j \equiv \gvec{x}_0$\;

\For{$t =  1, 2, \ldots$}{%
Receive new observation $\gvec{y}_t$\;

Sample $\{\dnnParams_j \}\stackrel{\rm i.i.d.}{\sim} q(\dnnParams|\bnnParams)$\;

\For{$j=1,2,\ldots,J$}{%
Predict $\hat{\gvec{x}}_{t|t-1}^j$ and  $\hat{\gvec{y}}_{t|t-1}^j$ from $\hat{\gvec{x}}_{t-1}$ via \eqref{eqn:ekf_pred}\;

Compute $\ac{kg}$ $\hat{\gvec{K}}^j_t$ via \ac{dnn} $\dnnParams_j$\;

Update $\hat{\gvec{x}}_{t}^j=\hat{\gvec{x}}_{t|t-1}^j + \hat{\gvec{K}}^j_t( {\gvec{y}}_{t} - \hat{\gvec{y}}_{t|t-1}^j)$

}
Compute state estimate $\hat{\gvec{x}}_{t}$ via \eqref{eqn:BKNMean}\;

Predict estimation error $\hat{\boldsymbol{\Sigma}}_{t}$ via \eqref{eqn:BKNVar}\;
}
}
\end{algorithm}
%
%%%%%%%%%%%%%%%%%%%%%%%%%%%%%%%%%%%
%%% Training Bayesian KalmanNet %%%
%%%%%%%%%%%%%%%%%%%%%%%%%%%%%%%%%%%
% \vspace{-0.1cm}
\subsection{Training Bayesian KalmanNet}
\label{ssec: Training Bayesian KalmanNet}
% \vspace{-0.1cm}
The trainable parameters of Bayesian KalmanNet are the distribution parameters $\bnnParams$. In principle, one can train Bayesian KalmanNet as a standard Bayesian regression \ac{dnn}, evaluating it based on the augmented \ac{ekf} output using the \ac{elbo} objective in \eqref{eq:ELBO}. Nonetheless, as we are particularly interested in accurate error covariance recovery (which is not explicitly accounted for in the \ac{elbo} formulation), we next propose a dedicated training loss, which is also geared towards our usage of Monte Carlo dropout for inducing stochasticity. 

We follow the specialization of the \ac{elbo} objective to concrete Monte Carlo based \acp{dnn} proposed in \cite{gal2017concretedropout}, while altering the data matching term to incorporate explicit encouraging of reliable error variance recovery, as  proposed in Subsection~\ref{ssec:Training} for frequentist learning. To formulate this, let $L$ denote the number of layers in the \ac{dnn} in which dropout is incorporated, and let $\{\lambda_l\}_{l=1}^L$ denote their indices. In the implementation detailed in Subsection~\ref{ssec: BKN Architectecture}, these indices correspond to the \ac{fc} layer of KalmanNet's Architecture 1. We write the parameters of layer $\lambda$ of $\dnnParams$ as $\dnnParams^\lambda$ and its dropout coefficient as $p^{(\lambda)}$. 
\textcolor{NewColor}{
Accordingly, the trainable parameters of Bayesian KalmanNet in such settings, denoted $\bnnParams$, include the weights of each of the layers $\dnnParams_0$, as well as the distribution parameters for the layers in which dropout is incorporated, denoted by $\pmb{p}=[p^{\lambda_1},\ldots,p^{\lambda_L}]$.}
The proposed loss function, based on the labeled data in \eqref{eq:dataset}, is formulated as  
\begin{align}
&\mySet{L}_{\mySet{D}}^{\rm BKN}(\bnnParams = [\dnnParams_0, \pmb{p}]) = \notag \\ 
&\mathbb{E}_{\dnnParams\sim q(\dnnParams|\bnnParams)}\{\mySet{L}_{\mySet{D}}^{\rm Emp}(\dnnParams)\}
+\sum_{l=1}^{L}\mySet{L}_{ \lambda_{l}}^{\rm KL}(\dnnParams_0^{\lambda_l}, p^{(\lambda_l)}),
\label{Loss BKN}
\end{align}
where $\mySet{L}_{ \lambda_{l}}^{\rm KL}(\bnnParams)$ computes the \ac{kl}-based regularization term of the layer $\lambda_{l}$ based on the learned multivariate Bernoulli distribution, with pre-determined prior in \eqref{eq:ELBO} being a maximal entropy Bernoulli distribution. \textcolor{NewColor}{The empirical loss term and the KL loss term in \eqref{Loss BKN} are balanced through the non-negative hyperparameters $c_1, c_2$, that appear in} the  layer-wise loss term, which is given by~\cite{gal2017concretedropout}
\begin{align}
     \mySet{L}_{ \lambda}^{\rm KL}(\dnnParams, p) = 
  c_{1}\cdot\frac{ \|\dnnParams\|^{2}}{1-p}-c_{2}\cdot \mySet{H}(p) \cdot n^{(\lambda)}.
     \label{Loss KL}
\end{align} 
In \eqref{Loss KL}, $n^{(\lambda)}$ is the number of input features to the $\lambda$th layer;  and $\mySet{H}(p)$ is the entropy of a Bernoulli random variable with probability $p$, defined as
\begin{align}
     \mySet{H}(p)=-p\log(p)-(1-p)\log(1-p).
     \label{Entropy}
\end{align}
This  loss in \eqref{Loss BKN} guides the tuning of $\bnnParams$ when training Bayesian KalmanNet to  both  extract an accurate state estimation and error covariance (by taking the expectation of our proposed empirical error variance as the data matching term), while encouraging randomness in the prior on the dropout probability $\pmb{p}$. \textcolor{NewColor}{Having formulated the loss in \eqref{Loss BKN}, the training of Bayesian KalmanNet follows the conventional stochastic gradient based optimization of \acp{dnn}, only with a Bayesian loss term. Accordingly,  the overall training complexity is comparable to that of frequentist KalmanNet~\cite{revach2022kalmannet}.}

% \begin{align}
%      \mySet{L}_{\mySet{D}}^{\rm \lambda_l}(\dnnParams)\!= 
%   c_{1}^{(l)}\cdot\frac{\sum_{j=1}^{N_{l}}\theta_{j}^{2}}{1-p^{(l)}}-c_{2}^{(l)}\cdot \mySet{H}(p^{(l)}) \cdot n^{(l)}.
%      \label{Loss KL2}
% \end{align}
% Where $N_{l}$ is the number of weights in layer $l$ and $n^{(l)}$ is the number of elements in the input to layer $l$. $c_{1}^{(l)}$ and $c_{2}^{(l)}$ are layer $l$'th coefficients, $p^{(l)}$ is layer l's dropout probability and $\mySet{H}(p)$ is the entropy of a Bernoulli random variable with probability $p$, defined as:
% \begin{align}
%      \mySet{H}(p)=-p\log(p)-(1-p)\log(1-p).
%      \label{Entropy}
% \end{align}
% This formulation allows the \ac{bkn} to train both to extract an accurate state estimation and covariance, while optimizing the dropout probability $p^{(l)}$ of each Bayesian layer.

%\vspace{-0.1cm}
\begin{table}[]
\centering
\caption{Domain knowledge requirements comparison.}
\begin{adjustbox}{width=\columnwidth} 
\begin{tabular}{|l|l|l|l|l|l|}
\hline
Method                                                                                                   & Feature   & $\gvec{F}(\cdot)$       & $\gvec{H}(\cdot)$       & $\gvec{R}$    & $\gvec{Q}$    \\ \hline\hline   
 \multirow{2}{*}{\begin{tabular}[c]{@{}l@{}}Output features\\ e.g., RKN\end{tabular}}   & State     & None\cellcolor[HTML]{AAFDB4} & None\cellcolor[HTML]{AAFDB4}    & None\cellcolor[HTML]{AAFDB4} & None\cellcolor[HTML]{AAFDB4} \\ \cline{2-6} 
& Error cov & None\cellcolor[HTML]{AAFDB4}    & None\cellcolor[HTML]{AAFDB4}    & None\cellcolor[HTML]{AAFDB4} & None\cellcolor[HTML]{AAFDB4} \\ \specialrule{.2em}{.1em}{.1em}
 \multirow{2}{*}{\begin{tabular}[c]{@{}l@{}}{\bf Bayesian KalmanNet}\end{tabular}}   & State     & Partial\cellcolor[HTML]{FFEAAD} & Partial\cellcolor[HTML]{FFEAAD}    & None\cellcolor[HTML]{AAFDB4} & None\cellcolor[HTML]{AAFDB4} \\ \cline{2-6} 
& Error cov & Partial\cellcolor[HTML]{FFEAAD}    & Partial\cellcolor[HTML]{FFEAAD}    & None\cellcolor[HTML]{AAFDB4} & None\cellcolor[HTML]{AAFDB4} \\ \specialrule{.2em}{.1em}{.1em}
\multirow{2}{*}{\begin{tabular}[c]{@{}l@{}}Learn KG and Prior\\ e.g., \ac{skn}\end{tabular}} & State     & Partial\cellcolor[HTML]{FFEAAD} & Partial\cellcolor[HTML]{FFEAAD} & None\cellcolor[HTML]{AAFDB4} & None\cellcolor[HTML]{AAFDB4} \\ \cline{2-6} 
& Error cov & Partial\cellcolor[HTML]{FFEAAD} & Full\cellcolor[HTML]{FF9595}    & None\cellcolor[HTML]{AAFDB4} & None\cellcolor[HTML]{AAFDB4} \\ \specialrule{.2em}{.1em}{.1em}
\multirow{2}{*}{\begin{tabular}[c]{@{}l@{}}Learn KG\\ e.g., KalmanNet\end{tabular}}                      & State     & Partial\cellcolor[HTML]{FFEAAD} & Partial\cellcolor[HTML]{FFEAAD} & None\cellcolor[HTML]{AAFDB4} & None\cellcolor[HTML]{AAFDB4} \\ \cline{2-6} 
 & Error cov & Partial\cellcolor[HTML]{FFEAAD} & Full\cellcolor[HTML]{FF9595}    & Full\cellcolor[HTML]{FF9595} & None\cellcolor[HTML]{AAFDB4} \\ \specialrule{.2em}{.1em}{.1em}
\multirow{2}{*}{\begin{tabular}[c]{@{}l@{}}Model-based\\ e.g., EKF\end{tabular}}                         & State     & Full\cellcolor[HTML]{FF9595}    & Full\cellcolor[HTML]{FF9595}    & Full\cellcolor[HTML]{FF9595} & Full\cellcolor[HTML]{FF9595} \\ \cline{2-6} 
& Error cov & Full\cellcolor[HTML]{FF9595}    & Full\cellcolor[HTML]{FF9595}    & Full\cellcolor[HTML]{FF9595} & Full\cellcolor[HTML]{FF9595} \\ \hline     
\end{tabular}
\label{tbl:QualtComp}
\end{adjustbox}
%\vspace{-0.3cm}
\end{table}

% \vspace{-0.1cm}
\subsection{Discussion}
\label{ssec: reduced domain knowledge}
% \vspace{-0.1cm}
Bayesian KalmanNet is designed to allow \ac{dnn}-aided tracking in partially known \ac{ss} models with error prediction. The extensions to existing frequentist KalmanNet-type methods proposed in Section~\ref{sec:DeepFrequent}, which map the interpretable internal features into the error covariance, follow the mathematical steps used by the \ac{ekf} in such computations, and thus necessitate additional domain knowledge not required by KalmanNet for state tracking.  Bayesian KalmanNet bypasses the need for such excessive knowledge by adopting \ac{ss} model-agnostic Bayesian deep learning techniques for error covariance extraction, combined with the \ac{ss} model-aware KalmanNet for state tracking. This results in Bayesian KalmanNet requiring the same level of partial \ac{ss} model knowledge as needed by KalmanNet and \ac{skn} for state estimation, while also providing error prediction, thus meeting \ref{itm:FuncAccuracy}-\ref{itm:Error}. As numerically reported in Section~\ref{sec:results}, Bayesian KalmanNet achieves reliable state tracking and error prediction in various settings with different levels of domain knowledge.

\textcolor{NewColor}{
 The uncertainty measure extracted by Bayesian KalmanNet encapsulates the overall estimation error covariance. This uncertainty measure inherently comprises two components~\cite{gawlikowski2023survey}: {\em aleatoric uncertainty}, resulting from the stochasticity of the system (e.g., process and measurement noises); and {\em epistemic uncertainty}, arising due to limited or mismatched knowledge within the estimator, particularly when the underlying dynamics are only partially known or when trained with a limited dataset. Classical model-based methods, such as the \ac{ekf}, can reliably quantify aleatoric uncertainty when operating with an accurate \ac{ss} model and well-characterized noise statistics, they underperform when the model is misspecified. Conversely, existing frequentist \ac{dnn}-aided \acp{kf} struggle to provide any form of uncertainty quantification, even with abundant training data. The approaches proposed for extending frequentist \ac{dnn}-aided \acp{kf} introduced in Section~\ref{sec:DeepFrequent} are geared towards quantifying aleatoric uncertainty, aiming to recreate the covariance computations of the model-based \ac{ekf} without explicitly propagating second-order moments. Accordingly, the resulting characterization, which does not encompass epistemic uncertainty, is expected to be sensitive to the amount of training data.  
 Bayesian KalmanNet overcomes these limitations by providing uncertainty estimates that reflect both sources of uncertainty, thereby yielding more reliable and calibrated error covariances under data scarcity and model mismatch. This dual capability makes Bayesian KalmanNet particularly suitable for practical scenarios where both the dynamics and noise statistics are only partially known.
}

The incorporation of Bayesian deep learning techniques can be viewed as trading computational complexity for uncertainty measures. This induced complexity also translates into additional inference latency, as reported in Section~\ref{sec:results}. \textcolor{NewColor}{The hyperparameter $J$ plays a key role in affecting both the performance, as increasing $J$ enhances the approximation of the posterior predictive distribution, while also increasing complexity. Specifically} the application of Bayesian KalmanNet can be viewed as $J$ realizations of KalmanNet with different parameterizations, thus increasing the computational burden by a factor of $J$ at each time instance. \textcolor{NewColor}{The induced latency of Bayesian KalmanNet can be notably reduced using parallel computing, as each of the $J$ realizations can be applied simultaneously.} We also note that this operation of Bayesian KalmanNet bears similarity to alternative  tracking algorithm that implement multiple Kalman-type filters in parallel, e.g., \cite{lefkopoulos2020interaction}, though the conventional approach uses a different \ac{ss} model with deterministic parameters for each filter, while Bayesian KalmanNet uses multiple stochastic realizations of the filter obtained for the same (partially known) \ac{ss} model, following the task-oriented discriminative learning paradigm~\cite{shlezinger2022discriminative}.

Our design of Bayesian KalmanNet gives rise to numerous possible avenues for extension. 
\textcolor{NewColor}{Our design, as well as that considered for frequentist KalmanNet~\cite{revach2022kalmannet}, assumes that the \ac{ss} model in \eqref{eq:gen_state} does not change. Thus, our approach is not naturally transferable to time-varying dynamics, for which adaptive \acp{kf}, often based on variational inference~\cite{huang2017novel,lan2023variational,lan2024joint}, are required. Previous works that considered the KalmanNet methodology in time-varying dynamics, which proposed adaptation based on hypernetworks~\cite{ni2024adaptive} and unsupervised learning~\cite{revach2022unsupervised}, focused on block-wise variations in the \ac{ss} model. The extension of the KalmanNet methodology to time-varying dynamics is the area of ongoing research, which, inspired by recent advances in continual Bayesian learning~\cite{jones2024bayesian, gusakov2025rapid}, can potentially benefit from the incorporation of Bayesian deep learning. Moreover,}
 while our formulation accommodates a generic Bayesian \ac{dnn}, our implementation \textcolor{NewColor}{detailed in Section~\ref{sec:results} is based on an architecture proposed in \cite{revach2022kalmannet},} and training objective focus on the usage of Monte Carlo dropout. Our formulation differs than standard Bayesian \acp{dnn} as all realizations are based on the same prior. One can possibly consider alternative Bayesian parameterizations, based on, e.g., variational inference with explicit priors~\cite{fortuin2022priors}, as well as its incorporation with different KalmanNet-type architectures, e.g.,~\cite{choi2023split,ni2022rtsnet,wang2024nonlinear,song2024practical}.
 Moreover, the combination of Bayesian deep learning and KalmanNet indicates that such methodologies can also be employed in alternative forms of model-based deep learning~\cite{shlezinger2023model}. We leave these extensions for future work.

%%%%%%%%%%%%%%%%%%%%%%%%%%%%%%%%%%%%%%%%%%%%%%%%%%%%%%%%%%%%%%%%%%%%%%%%%%%%%%%%
% \vspace{-0.2cm}
\section{Numerical Evaluation}
\label{sec:results}
% \vspace{-0.1cm}
% \textbf{This is the easiest part for you to write}
% \textbf{Start by explaining the considered tracking algorithm}
% \textbf{Then present the synethetic \ac{ss} model and discuss the results.}
In this section we numerically compare the above techniques to extract error covariance in \ac{dnn}-aided \acp{kf}. We first detail the considered algorithms implemented and our utilized performance measures in Subsection~\ref{ssec:results_Algo}. Then, we detail a set of experiments dedicated to evaluating state and covariance estimation in different \ac{ss} models in Subsections~\ref{ssec:ExpCanonical}-\ref{ssec:CV SS}. We conclude by comparing inference time in Subsection~\ref{ssec:results_Inf}.
%
%%%%%%%%%%%%%%%%%%%%%%%%%%
%%% Experimental Setup %%%
%%%%%%%%%%%%%%%%%%%%%%%%%%
%
\subsection{Experimental Setup}
\label{ssec:results_Algo}
\subsubsection{Tracking Algorithms}
We henceforth compare the following tracking algorithms\footnote{The source code and all hyperparameters used can be found online at \url{https://github.com/yonatandn/Uncertainty-extraction-in-Model-Based-DL}}:
\begin{itemize}
    \item {\bf \ac{rkn}}, using the code provided in \cite{becker2019recurrent}, representing \acp{dnn} with dedicated output features.
    \item {\bf \ac{skn}}, using the code provided in \cite{choi2023split}, representing model-based deep learning  with learned \ac{kg} and prior covariance.
    \item KalmanNet, using both Architecture 1 ({\bf KNetV1}) and  Architecture 2 ({\bf KNetV2}) of \cite{revach2022kalmannet}.
    \item Bayesian KalmanNet (termed {\bf BKN} for brevity), based on KNetV1 with dropout incorporated into its \ac{fc} layers, taking $J=20$ realizations.
    \item The fully model-based {\bf \ac{ekf}}.
\end{itemize}
%For \acp{dnn} with dedicated output features, we employ \ac{rkn} using the code provided in \cite{becker2019recurrent}. For model-based deep learning methods computing the \ac{kg} we use \ac{skn}~\cite{choi2023split} (coined {\em Split KN}), KalmanNet~\cite{revach2022kalmannet}, where the latter does not compute the prior covariance and the \ac{bkn} architecture presented in \ref{sec:BKN}. 
Unless stated otherwise, we use \eqref{eqn:ErrorCovSKN} to convert the \ac{kg} and prior covariance into error covariance. We train \ac{rkn} and \ac{skn} with \eqref{eqn:GNLL_Loss}; KalmanNet is trained with \eqref{eqn:Loss}; while Bayesian KalmanNet is trained using \eqref{Loss BKN}. All data-driven algorithms are trained with the same data comprising of $|\mySet{D}|=300$ trajectories. \textcolor{NewColor2}{ Training commences with the \ac{dnn} weights initialized via standard Xavier initialization, and the Monte Carlo dropout parameters of BKN randomized uniformly in $[0.5,0.8]$, following the common practice in Bayesian \acp{dnn}~\cite{gal2017concretedropout}}. The remaining learning hyperparameters were selected based on standard cross validation. 

\subsubsection{Data generation}
\label{ssec:results_Est}
%We numerically compare the above techniques to extract error covariance in \ac{dnn}-aided \acp{kf}.  
Unless stated otherwise, throughout the following experimental study, we set the observation noise covariance $\gvec{R}$ to take the form $r^2 \gvec{I}$, with $1/r^2 = n/{\rm Tr}(\gvec{R})$ denoting the \ac{snr}. The ratio between the state evolution and the observation noise variances is fixed to $0.01$, such that changes in the \ac{snr} affect both the state evolution and the observation noises. \textcolor{NewColor}{All \ac{dnn}-aided algorithms were trained on a range of \acp{snr} rather than for each \ac{snr} separately, to emphasize their robustness and ability to operate across different noise levels.} To perform  statistical tests and evaluation, \acl{mc} simulation was performed, with $N=100$ trajectories per scenario.

\subsubsection{Performance Measures}
The tracking algorithms are evaluated in terms of the following performance measures:
$(i)$ The empirical estimation error, i.e., state estimate \ac{mse} \eqref{eq:MSE}; and
$(ii)$  The predicted error covariance, namely, ${\rm Tr}( \hat{\boldsymbol{\Sigma}}_t)$.
Performance is thus measured not only in state estimation error, but also on the ability to predict this error from the recovered covariance. 

To assess the credibility of the estimators, we also evaluate the \ac{anees} measure~\cite{BARSHALOM1983431, bar2004estimation, Drummond1998ComparisonOV,Credibility}. This measure combines state estimation and error covariance prediction in a single measure. For an estimator outputting state estimates and covariances $\{\hat{\gvec{x}}_t, \hat{\boldsymbol{\Sigma}}_t\}_{t=1}^T$, the \ac{anees} is given by
\begin{align}
     {\rm ANEES} =  \frac{1}{T} \sum_{t=1}^T 
     \mathbb{E}\left\{(\gvec{x}_t - \hat{\gvec{x}}_t)^\top\hat{\boldsymbol{\Sigma}}_t^{-1}(\gvec{x}_t - \hat{\gvec{x}}_t)  \right\},
     \label{ANEES}
\end{align}
where the stochastic expectation is computed by empirical averaging over the test data. 
The closer to the \ac{anees} is to being $1$, the more credible the estimator, with values surpassing $1$ representing overconfidence, and lower values corresponding to conservative predictions.

Additional metrics used to evaluate the different methods at a given time instance $t$ over a test set $\mySet{D}$  are the \ac{apec} and the \ac{eec}.
The \ac{apec} is given by
\begin{align}
     {\rm \ac{apec}}_t(\mySet{D}) =  \frac{1}{|\mySet{D}|} \sum_{i=1}^{|\mySet{D}|} 
     \hat{\boldsymbol{\Sigma}}^{(i)}_t,
     \label{Average Predicted Covariance}
\end{align}
%where \(N\) is the number of \acl{mc} simulations, and superscript \((n)\) is the \(n\)'th run.
%
while the \ac{eec} is given by
\begin{align}
     {\rm \ac{eec}}_t(\mySet{D})  =  \frac{1}{|\mySet{D}|} \sum_{i=1}^{|\mySet{D}|} 
     (\gvec{x}_t - \hat{\gvec{x}}^{(i)}_t)(\gvec{x}_t - \hat{\gvec{x}}^{(i)}_t)^\top.
     \label{Empirical Error Covariance}
\end{align}
We say that the algorithm is a better uncertainty estimator as its uncertainty prediction (\ac{apec} along time) is closer to the true error covariance (\ac{eec} along time).

\subsubsection{Mismatch Setup}
We consider the following forms of mismatches in the model knowledge:
\begin{itemize}
    \item {\em Process noise mismatch}: a dataset generated with a process noise covariance matrix $\gvec{Q}_{\rm data}=100\cdot\gvec{Q}_{\rm model}$ where $\gvec{Q}_{\rm model}$ is the process noise covariance matrix known to the algorithm.
    \item {\em Measurement noise mismatch}: a dataset generated with a measurement noise covariance matrix $\gvec{R}_{\rm data}=100\cdot\gvec{R}_{\rm model}$ where $\gvec{R}_{\rm model}$ is the measurement noise covariance matrix known to the algorithm.
    \item {\em Model mismatch}: in the \ac{cv} \ac{ss}, data was generated using a \ac{ca} model, while testing the different methods with a \ac{cv} \ac{ss}.
\end{itemize}
\textcolor{NewColor}{Our motivation for employing the considered mismatches, imposed on all algorithms that require knowledge of the \ac{ss} model, is to evaluate  sensitivity to incorrect modeling assumptions. Specifically, the noise mismatches (which affect just the \ac{ekf}) serve to highlight the behavior of uncertainty quantification under severe mismatches, while the last model mismatch represents a realistic setup in which a target follows a more complicated state evolution model compared to the one used by the tracking algorithm.}
\subsection{Linear Canonical SS Model}
\label{ssec:ExpCanonical}
We first simulate a synthetic two-dimensional linear \ac{ss} model with Gaussian process and measurement noises governed by a controllable canonical state transition matrix, defined in \cite[Sec. IV.B]{revach2022kalmannet}. Here, the state transition matrix \( {\gvec{F}} \) and measurement matrix \( {\gvec{H}} \) are
\[
% \gvec{x}(t) = 
% \begin{bmatrix} x_1(t) \\ x_2(t) \end{bmatrix} , \quad 
{\gvec{F}} = \frac{1}{\sigma_{\text{max}}} \begin{bmatrix}
1 & 1\\
0 & 1
\end{bmatrix} , \quad 
{\gvec{H}} = \begin{bmatrix} 1 & 0 \\ 0 & 1 \end{bmatrix},
\]
where \( \sigma_{\text{max}} \) is the largest singular value of \( \hat{\gvec{F}}_t \) ensuring stability.

\begin{figure*}
\begin{subfigure}{1\columnwidth}
\includegraphics[width=1\linewidth]{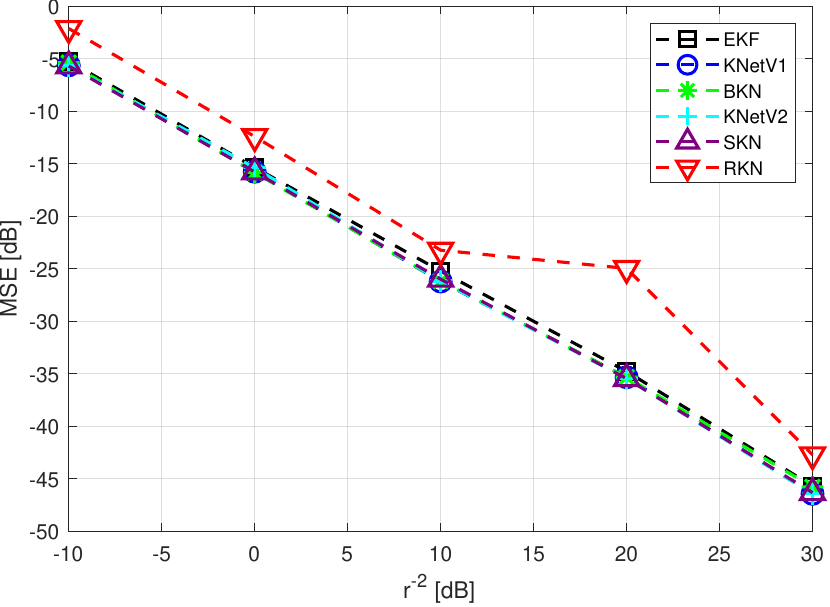}
\caption{\ac{mse} for mismatched $\gvec{Q}$, all the model aware methods managed to track the state with a very similar \ac{mse}. The model agnostic \ac{rkn} has failed to maintain the same performance.}
\label{fig:Canonical Mismatch Q MSE}
\end{subfigure}
%
%%%%% Mismatch in Q
\begin{subfigure}{1\columnwidth}
\includegraphics[width=\linewidth]{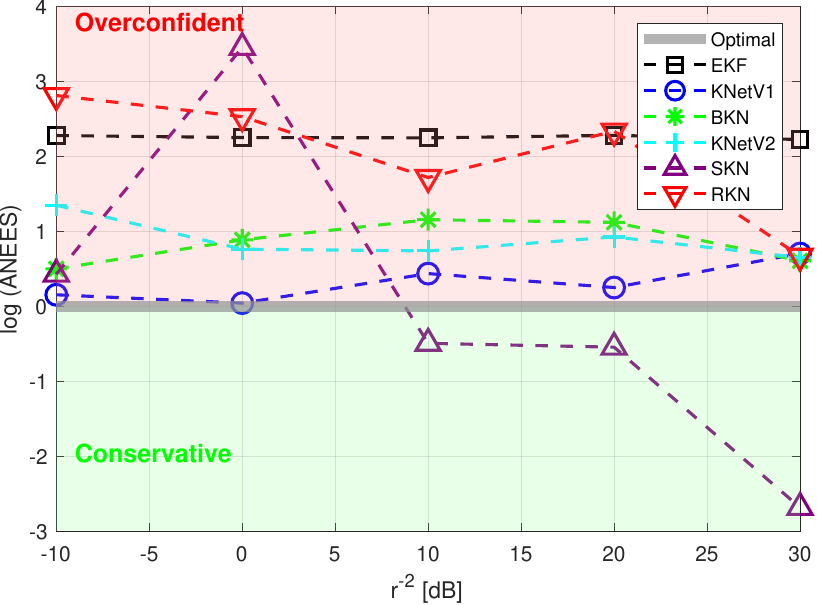}
\caption{\ac{anees} for mismatched $\gvec{Q}$, Both the frequentist and Bayesian KNet have managed to mainatain a stable error covariance (uncertainty) estimation throughout the different \acp{snr}.% - indicated by the proximity to the optimal $\log(\ac{anees})=0$. As can also be seen, the \ac{skn}'s error covariance prediction were unstable, and both the \ac{ekf} and the \ac{rkn} have predicted a very overconfident estimation of the error covariance
}
\label{fig:Canonical Mismatch Q ANEES}
\end{subfigure}
%
%%%%%%%%%%%%%%
%%% Figure %%%
%%%%%%%%%%%%%%
%
\begin{subfigure}{1\columnwidth}
\includegraphics[width=\columnwidth]{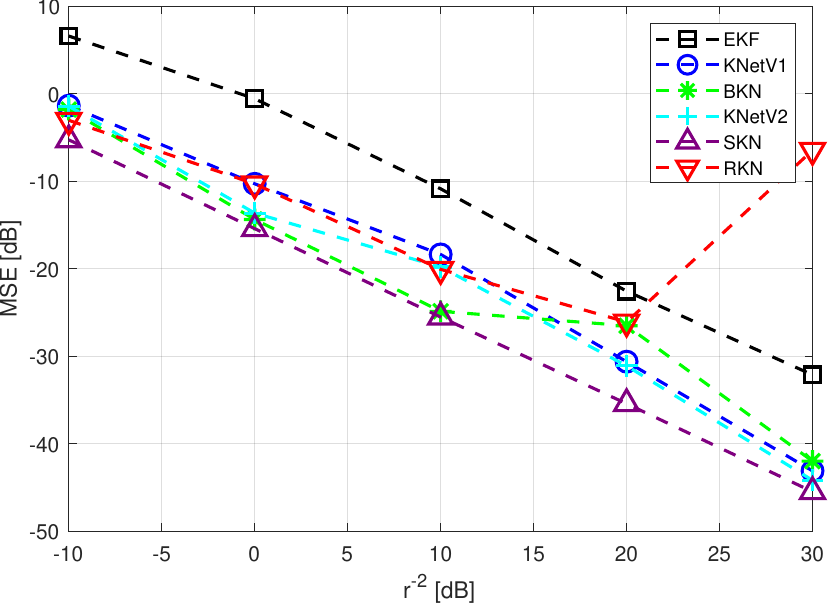}
\caption{\ac{mse} for mismatched $\gvec{R}$, the \ac{skn} performance was superior while both the frequentist and Bayesian model-based deep learning methods managed to track the state with a very similar \ac{mse}. %, whereas and the model based \ac{ekf} achieved a poor state prediction.
}
\label{fig:Canonical Mismatch R MSE}
\end{subfigure}
%
%%%%% Mismatch in R
\begin{subfigure}{1\columnwidth}
\includegraphics[width=\columnwidth]{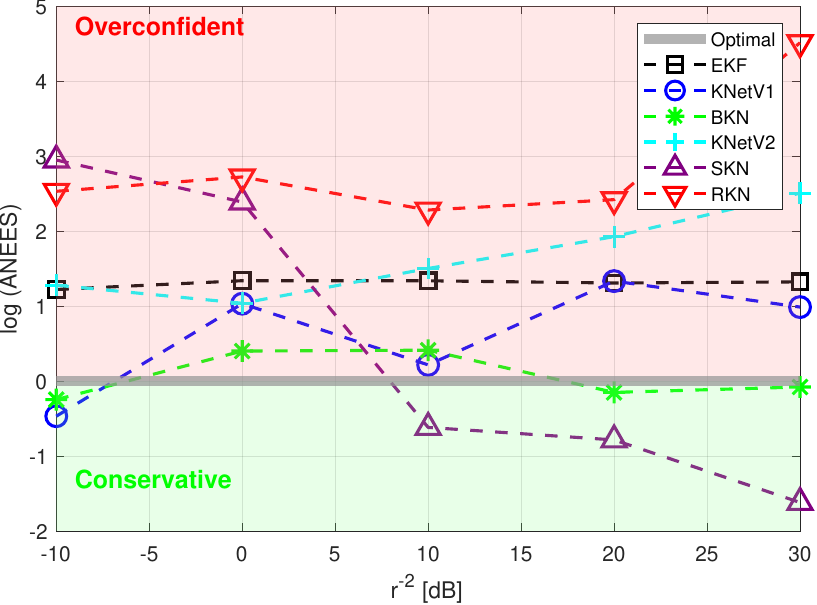}
\caption{\ac{anees} for mismatched $\gvec{R}$, the \acl{bkn} consistently provided the most accurate and stable uncertainty estimates, whereas the model agnostic \ac{rkn}'s prediction was the mostly inaccurate.
%, as shown by its proximity to the optimal $\log (\ac{anees}) = 0$ across all \acp{snr}. In contrast, the RKN method was the worst, predicting a highly overconfident error covariance. The \ac{ekf} showed better performance than KNetV2 in some \acp{snr} ranges.
}
\label{fig:Canonical Mismatch R ANEES}
\end{subfigure}
\caption{Canonical linear \ac{ss} model - mismatched process noise $\gvec{Q}$ and measurement noise $\gvec{R}$ covariances}
\end{figure*}

The considered performance measures versus the observations noise variance $r^2$ are reported in Fig.~\ref{fig:Canonical Mismatch Q ANEES}. 
 When examining the \ac{mse} in Fig.~\ref{fig:Canonical Mismatch Q MSE}, we can see that all the model aware methods achieved a smaller estimation error compared to the \ac{rkn} which is model agnostic.
In Fig.~\ref{fig:Canonical Mismatch Q ANEES} we observe that (frequentist) KalmanNet provides the most accurate error covariance estimates compared to the other methods when introducing a mismatch in the process noise covariance $\gvec{Q}$. This is evidenced by the lowest absolute value of $\log(\ac{anees})$, suggesting that KalmanNet closely matches the true uncertainty in the system.  This stems from the fact that in this relatively simple setup, one can translate the \ac{kg} into an error covariance estimate as detailed in Section~\ref{sec:DeepFrequent}. 
% ($log(\ac{anees}) = 0$, indicating perfect alignment between predicted and actual uncertainty, reflecting the filter’s credibility)
The \ac{skn} produces either too conservative or too confident predictions, meaning it brutally overestimates or underestimates the uncertainty, leading to larger error covariances.   \acl{bkn} demonstrates minor overconfidence in its predictions, underestimating the uncertainty. Both  \ac{rkn} and \ac{ekf} exhibit significant overconfidence, with their predictions being excessively optimistic. This means their error covariances are underestimated and are not aligned with the true uncertainties.

When introducing a mismatch in the measurement noise covariance $\gvec{R}$ and evaluating the filter performance using the \ac{anees} metric, as shown in Fig.~\ref{fig:Canonical Mismatch R ANEES}, it is evident that  \acl{bkn} achieves the most favorable results. \acl{bkn} provides a credible error covariance estimation, meaning it maintained reliability. The \ac{ekf} and KalmanNet produce comparable results, indicating that both methods offer similar levels of credibility in their error covariance predictions. In contrast, the \ac{rkn} struggles to provide credible estimates, leading to unreliable uncertainty predictions.
\textcolor{NewColor}{Comparing Figs.~\ref{fig:Canonical Mismatch Q ANEES} and~\ref{fig:Canonical Mismatch R ANEES} shows that  frequentist methods can also achieve strong performance, particularly in scenarios where uncertainty can be reliably estimated. However, in cases where uncertainty estimation is inherently more challenging, e.g., where inaccuracies directly impact the observed data—frequentist methods relying on internal features may struggle. In such settings, Bayesian KalmanNet provides more robust and calibrated uncertainty estimates by explicitly modeling parameter uncertainty through ensemble predictions. Overall, both variants exhibit stable and credible uncertainty estimates across different SNR levels, and the observed differences are relatively minor.}

Turning to the \ac{mse} results, illustrated in Fig.~\ref{fig:Canonical Mismatch R MSE}, it is clear that the \ac{skn}, \textcolor{NewColor}{which is particularly designed for settings with imbalances between state evolution and measurement noises~\cite{choi2023split},} outperforms other methods by achieving the lowest estimation error across various measurement noise covariance levels. This indicates that \ac{skn} consistently provided the most accurate state estimates, regardless of the noise magnitude. The \ac{rkn}, however, only performed well at higher $r^2$ values, suggesting that it struggles in high \acp{snr} \textcolor{NewColor}{ with limited data sets,} where accurate uncertainty modeling is more critical. All KalmanNet based methods, including \acl{bkn}, outperform the \ac{ekf}. \textcolor{NewColor}{Comparing the \ac{mse} results in Fig.~\ref{fig:Canonical Mismatch R MSE} and the uncertainty results in Fig.~\ref{fig:Canonical Mismatch R ANEES} indicates on the trade-off emerging by the need to balance  both objectives during training. In such multi-objective settings, improving one aspect-—e.g., achieving lower MSE-—may slightly compromise the precision of uncertainty estimates, sometimes somewhat compromise the precision of uncertainty estimates, and in some cases hardly affect the latter. }

The results in this \ac{ss} model emphasize that for a variety of \acp{snr}, even for a linear state transition and measurement functions, both the frequentist and the Bayesian methods combining model knowledge and deep learning achieve better results than both model agnostic deep learning and model based without deep learning.
%
%%%%%%%%%%%%%%%%%%%%%%%%%%%%%%%%%
%%% Figure: Canonical \ac{ss} %%%
%%%%%%%%%%%%%%%%%%%%%%%%%%%%%%%%%
%
%
%%%%%%%%%%%%%%%%%%%%%%%
%%%%%% PENDULUM %%%%%%%
%%%%%%%%%%%%%%%%%%%%%%%
%
\subsection{Pendulum SS Model}
\label{ssec:Pendulum}
Next, we consider a pendulum with mass \( m=1 \) and length \( \ell=1 \). The pendulum exhibits nonlinear behavior due to the gravitational force \( g \) acting on it. We describe the system using the angle \( \theta \) and angular velocity \( \dot{\theta} \) as the state variables, sampled at interval of $\Delta t = 0.01$.
Let the state vector \( \gvec{x}_t\) be defined as:
\[
\gvec{x}_t = \begin{bmatrix} \theta_t \\ \dot{\theta}_t \end{bmatrix}.
\]
%where \( \theta(t) \) is the angular displacement and \( \dot{\theta}(t) \) is the angular velocity.
The dynamics of the pendulum can be modeled using Newton’s second law. For a pendulum subject to gravity, 
%the nonlinear differential equation is:
% \[
% \ddot{\theta}(t) + \frac{g}{\ell} \sin(\theta(t)) = 0
% \]
%
we can express this second-order equation as a system of first-order differential equations
% :
% \[
% \dot{\gvec{x}}(t) = \begin{bmatrix} \dot{\theta}(t) \\ -\frac{g}{\ell} \sin(\theta(t)) \end{bmatrix}
% \]
such that  
%  the \ac{ss} representation becomes:
% \[
% \dot{\gvec{x}}(t) = \begin{bmatrix} x_2(t) \\ -\frac{g}{\ell} \sin(x_1(t)) \end{bmatrix}
% \]
% Using numerical integration with a discrete time step \( \Delta t \), 
the discrete nonlinear state transition function \( \gvec{F}(\gvec{x}_t) \) can be written as:
\[
\gvec{f}(\gvec{x}_t) = \begin{bmatrix} \theta_t + \dot{\theta}_t \Delta t \\ \dot{\theta}_t - \frac{g}{\ell} \sin(\theta_t) \Delta t \end{bmatrix}.
\]
%
%The state transition matrix \( \hat{\gvec{F}}_{t} \) can be derived from the nonlinear equations simply by calculating the jacobian of $\gvec{F}(\gvec{x}_t)$. 
Assuming that \( \gvec{x}_t \) at each time step is small enough to allow a local linearization, the Jacobian matrix of $\gvec{F}(\gvec{x}_t)$ is 
\[
\hat{\gvec{F}}_{t} = \begin{bmatrix} 1 & \Delta t \\ -\frac{g}{\ell} \cos(\theta_t) \Delta t & 1 \end{bmatrix}.
\]

We assume the pendulum’s position in Cartesian coordinates, 
% \( (x(t), y(t)) \) can be observed, where:,\
% %
% \[
% x(t) = \ell \sin(\theta(t)), \quad 
% y(t) = -\ell \cos(\theta(t))
% \]
% %
% Thus 
such that
the observation matrix \( {\gvec{h}}(\gvec{x}_t) \) for converting state to measurement is given by:
\[
{\gvec{h}}(\gvec{x}_t)  = 
\begin{bmatrix}
\ell\cos(\theta_t) \\ 
\ell\sin(\theta_t)
\end{bmatrix}.
\]
Here, \( \hat{\gvec{H}}(\gvec{x}_t) \) is non-full columns rank. This means that the frequentist model-based methods are unable to extract an error covariance matrix as explained in Remark~\ref{rmk:Partial}, and thus are not  tested throughout this scenario.

\begin{figure*}
%
%%%%%%%%%%%%%%%%%%%%%%%%%%%%%%%%%%%%%%%
%%% Figure: Pendulum Mismatch Q MSE %%%
%%%%%%%%%%%%%%%%%%%%%%%%%%%%%%%%%%%%%%%
%
\begin{subfigure}{1\columnwidth}
\includegraphics[width=\linewidth]{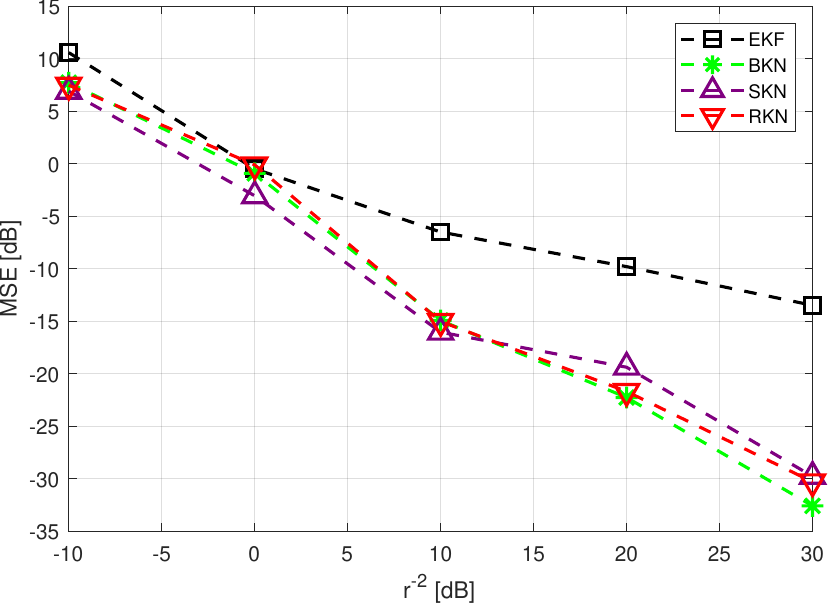}
%\vspace{-0.7cm}
\caption{\ac{mse} for mismatched $\gvec{Q}$, All the deep learning methods showed stable results but struggled slightly at moderate \acp{snr}. The \ac{ekf}, however, failed to maintain competitive performance.%, with significantly higher \ac{mse} across all \acp{snr}, indicating sensitivity to the process noise mismatch.
}
\label{fig:Pendulum Mismatch Q MSE}
%\vspace{-0.3cm}
\end{subfigure}
%
%%%%%%%%%%%%%%%%%%%%%%%%%%%%%%%%%%%%%%%%%
%%% Figure: Pendulum Mismatch Q ANEES %%%
%%%%%%%%%%%%%%%%%%%%%%%%%%%%%%%%%%%%%%%%%
%
\begin{subfigure}{1\columnwidth}
\centering
\includegraphics[width=\linewidth]{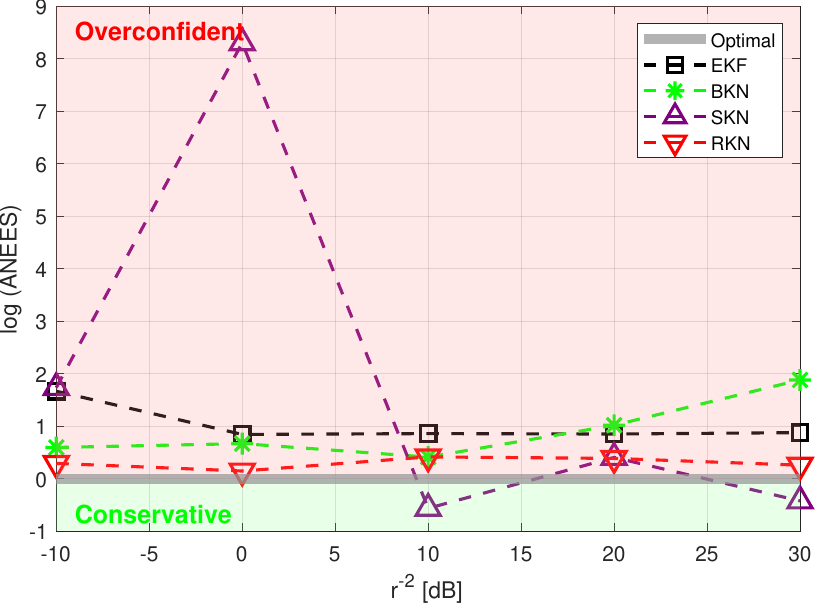}
%\vspace{-0.7cm}
\caption{\ac{anees} for mismatched $\gvec{Q}$, the different methods provided very similar results, beside the instability of the \ac{skn}'s credibility for some \acp{snr}.}
\label{fig:Pendulum Mismatch Q ANEES}
%\vspace{-0.3cm}
\end{subfigure}
%
%%%%%%%%%%%%%%%%%%%%%%%%%%%%%%%%%%%%%%%%
%%% Figure : Pendulum Mismatch R MSE %%%
%%%%%%%%%%%%%%%%%%%%%%%%%%%%%%%%%%%%%%%%
%
\begin{subfigure}{1\columnwidth}
\includegraphics[width=\linewidth]{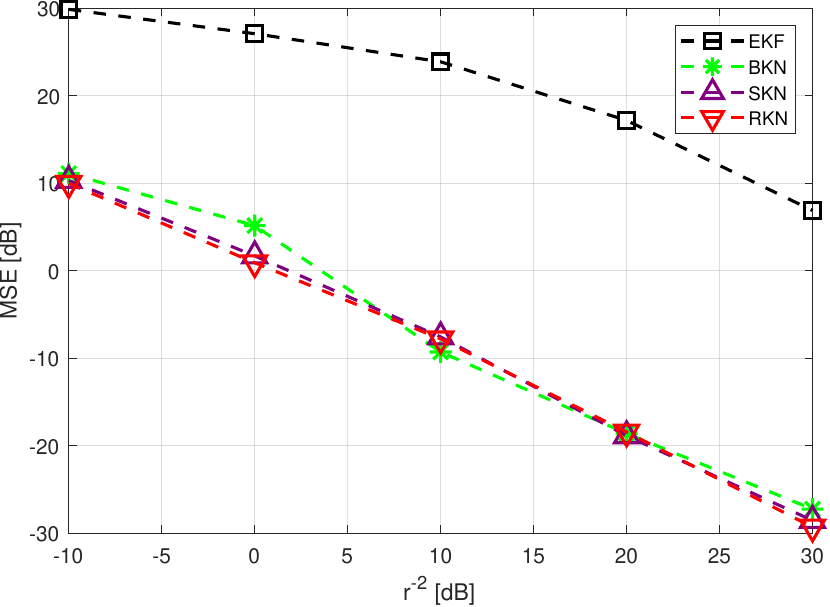}
%\vspace{-0.7cm}
\caption{\ac{mse} for mismatched $\gvec{R}$, All the deep learning methods showed stable results with a significantly better performance than the \ac{ekf} in the sense of the estimated state accuracy.
}
\label{fig:Pendulum Mismatch R MSE}
%\vspace{-0.3cm}
\end{subfigure}
%
%%%%%%%%%%%%%%%%%%%%%%%%%%%%%%%%%%%%%%%%%
%%% Figure: Pendulum Mismatch R ANEES %%%
%%%%%%%%%%%%%%%%%%%%%%%%%%%%%%%%%%%%%%%%%
%
\begin{subfigure}{1\columnwidth}
\includegraphics[width=\linewidth]{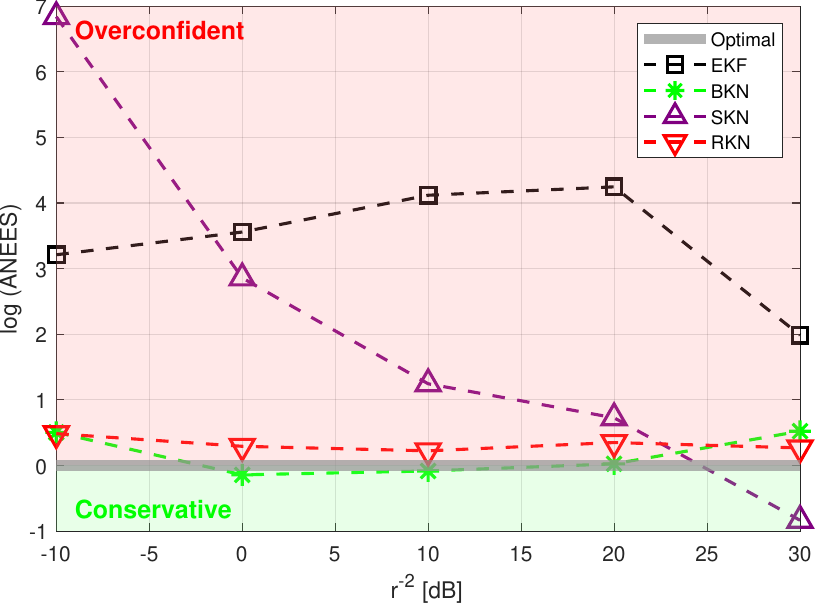}
%\vspace{-0.7cm}
\caption{\ac{anees} for mismatched $\gvec{R}$, the model based deep learning methods consistently provide the most accurate and stable uncertainty estimates compared to the model-free and model dependent methods.}
\label{fig:Pendulum Mismatch R ANEES}
%\vspace{-0.3cm}
\end{subfigure}
\caption{Pendulum non-linear \ac{ss} model - mismatched process noise $\gvec{Q}$ and measurement noise $\gvec{R}$ covariances}
\label{fig:Pendulum}
\end{figure*}

The achieved performance measures are reported in Fig.~\ref{fig:Pendulum}.
When examining the \ac{mse} in Fig.~\ref{fig:Pendulum Mismatch Q MSE}, it can be seen that the \ac{ekf} fails to produce an accurate state estimation compared to all other methods, especially at higher \acp{snr}.
The \ac{anees} results in Fig.~\ref{fig:Pendulum Mismatch Q ANEES} highlight that beside the \ac{skn}, all algorithms have a stable credibility with a minor overconfidence throughout different \acp{snr}, whereas the \ac{skn} shows some instability in its credibility. The \ac{rkn}  provides the most accurate error covariance estimates compared to the other methods when introducing a mismatch in the process noise covariance $\gvec{Q}$.

Under mismatched $\gvec{R}$, the \ac{mse} results in Fig.~\ref{fig:Pendulum Mismatch R MSE} show that in terms of estimation error the \ac{ekf} performs the worst, as its MSE remains significantly higher than the other methods across the entire range of $1/r^2$. This indicates that the \ac{ekf} struggles to maintain accurate state estimates, particularly in the presence of higher measurement noise. Its performance gap widens considerably as the noise mismatch increases, showing that the \ac{ekf} is less robust under such conditions. The \ac{dnn}-aided \acl{bkn}, \ac{skn},  and \ac{rkn} all perform much better, achieving comparable results for most of the range.

Fig.~\ref{fig:Pendulum Mismatch R ANEES} shows that the \ac{skn} is the most overconfident filter only at higher noise levels, though it improves compared to the other methods as the noise level is lower. The \ac{ekf}, demonstrates a large overconfidence, throughout all noise levels. In contrast, \acl{bkn} and the \ac{rkn} consistently yield a very accurate (credible) error covariance estimates as it stays closest to the optimal $\log(\ac{anees}) = 0$ for much of the range. 
\textcolor{NewColor}{To highlight the dual gains of Bayesian KalmanNet in accuracy and uncertainty, we also include a tabular comparison of the \ac{mse} and \ac{anees} of the considered algorithms for the setting with mismatched $\gvec{R}$ for fixed \ac{snr} of $1/r^2=10$ [dB]. The results, reported in Table~\ref{tbl:Metric for table in revision}, clearly indicate on the ability of Bayesian KalmanNet to simultaneously provide accurate tracking alongside faithful uncertainty estimation, notably outperforming both frequentist \ac{dnn}-based methods as well as the model-based \ac{ekf}.}

\begin{table}
\centering
\caption{Performance metrics at \ac{snr} $1/r^2 = 10$ [dB] with mismatched $\gvec{R}$, Pendulum non-linear \ac{ss} model.}
\label{tbl:Metric for table in revision}
\vspace{0.2cm}
%\begin{adjustbox}{width=\columnwidth} 
\begin{tabular}{|c|c|c|c|c|}
\hline
 {\bf{Metric}}  &   {\bf{EKF}}  &      {\bf{SKN}}  &   {\bf{RKN}}  &   {\bf{BKN}}    \\ \hline
 {MSE [dB]}  &  {23.8933}\cellcolor[HTML]{FF9595}   &     {-7.5536}  &   {-7.7836}  &   {-9.2741}\cellcolor[HTML]{AAFDB4}    \\ \hline % JobID 1013851
     {log(ANEES)}  & {4.1182}\cellcolor[HTML]{FF9595}  &     {1.2452}  &   {0.2234}  &   {-0.0834}\cellcolor[HTML]{AAFDB4}
   \\ \hline 
\end{tabular}
%\end{adjustbox}
\vspace{-0.3cm}
\end{table}

%
%%%%%%%%%%%%%
%%% Figure %%
%%%%%%%%%%%%%
%
%
%%%%%%%%%%%%%%%%%%%%%%%%%
%%%%%% NAVIGATION %%%%%%%
%%%%%%%%%%%%%%%%%%%%%%%%%
%
\subsection{Navigation SS Model}
\label{ssec:CV SS}
Our third scenario focuses on a more realistic \ac{ss} as we track the location and velocity of a vehicle moving in a 2D plane. We assume its movement is in a \ac{cv} manner. The system is modeled using the vehicle’s position and velocity in the \( x \)- and \( y \)-directions as state variables. 

Let the state vector \( \gvec{x}_t \) and state transition matrix \( {\gvec{F}} \) be defined as:
\[
\gvec{x}_t = 
\begin{bmatrix}
x_t \\ y_t \\ \dot{x}_t \\ \dot{y}_t 
\end{bmatrix}, \quad 
{\gvec{F}}= \begin{bmatrix} 
1 & 0 & \Delta t & 0 \\
0 & 1 & 0 & \Delta t \\
0 & 0 & 1 & 0 \\
0 & 0 & 0 & 1 
\end{bmatrix}
\]
where:
\begin{itemize}
\item \( x_t \) and \( y_t \) represent the position of the vehicle in the \( x \)- and \( y \)-directions.
\item \( \dot{x}_t \) and \( \dot{y}_t \) represent the velocity of the vehicle in the \( x \)- and \( y \)-directions.
\end{itemize}
%
% For a \ac{cv} motion, the state transition matrix in discrete form is:
% \[
% \hat{\gvec{F}}_t = \begin{bmatrix} 
% 1 & 0 & \Delta t & 0 \\
% 0 & 1 & 0 & \Delta t \\
% 0 & 0 & 1 & 0 \\
% 0 & 0 & 0 & 1 
% \end{bmatrix}
% \]

Assume we can observe the position of the vehicle in the \( x \)- and \( y \)-directions, but not its velocity. 
The observation matrix \( {\gvec{H}} \) maps the state vector to the measurements:
\[
{\gvec{H}} = 
\begin{bmatrix} 
1 & 0 & 0 & 0 \\
0 & 1 & 0 & 0 
\end{bmatrix}
\]
\textcolor{NewColor}{We generate a set of 100 trajectories, with 5 used for validation and 10 reserved for testing. The sampling interval parameter $\Delta t$ is normalized to unity.}

\begin{figure*}
\begin{subfigure}{1\columnwidth}
\includegraphics[width=\linewidth]{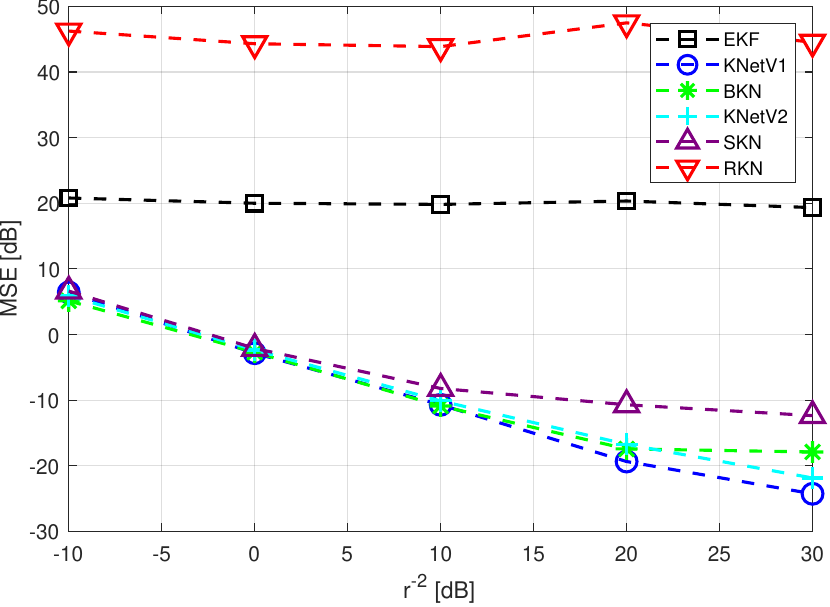}
%\vspace{-0.7cm}
\caption{\ac{mse} for mismatched $\gvec{Q}$, the model based deep learning methods consistently provide the most accurate state estimates compared to the model-free and model dependent methods.
}
\label{fig:CV Mismatch Q MSE}
%\vspace{-0.3cm}
\end{subfigure}
%%%%% Mismatch in Q
\begin{subfigure}{1\columnwidth}
\includegraphics[width=\linewidth]{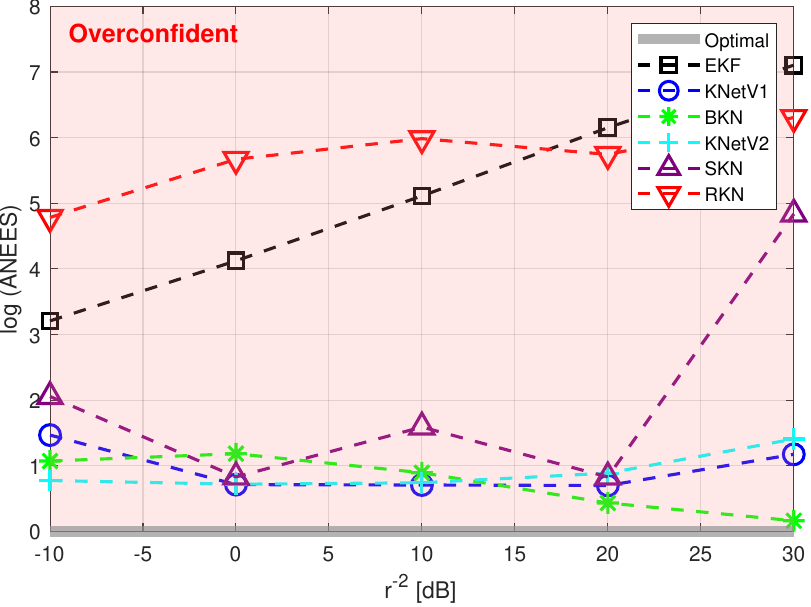}
%\vspace{-0.7cm}
\caption{\ac{anees} for mismatched $\gvec{Q}$, Both the frequentist and \acl{bkn}  manage to maintain an accurate error covariance estimation throughout the different \acp{snr}.
}
\label{fig:CV Mismatch Q ANEES}
%\vspace{-0.3cm}
\end{subfigure}
\caption{\ac{cv}  \ac{ss} model - mismatched process noise $\gvec{Q}$}
\label{fig:CV}
\end{figure*}

The results for this setting are reported in Fig.~\ref{fig:CV}. 
When examining the \ac{mse}   in Fig.~\ref{fig:CV Mismatch Q MSE}, it can be seen that both the \ac{ekf} and the \ac{rkn} not only fail to estimate the uncertainty, but also fail to produce an accurate state estimation compared to all other methods, especially at higher \acp{snr}.
The results in Fig.~\ref{fig:CV Mismatch Q ANEES} shows that all methods resulted in an overconfident covariance prediction, although the both the frequentist and Bayesian model based deep learning methods manage to have a much better credibility with a more accurate uncertainty predictions.

%
%%%%%%%%%%%%%%%%%%%%%%%%%%%%%%%%%%
%%% CV for mismatched $\gvec{Q} %%%
%%%%%%%%%%%%%%%%%%%%%%%%%%%%%%%%%%
%
In this section, we describe a single run of our constant velocity model under specific conditions. The true process noise covariance \( \gvec{Q}_{\rm data} \) was set such that the given process noise covariance \( \gvec{Q}_{\rm model} \) is underestimated by a factor of 100, i.e., \( \gvec{Q}_{\rm data} = 100\cdot \gvec{Q}_{\rm model} \). The noise covariance \( \gvec{R} \) was set to $10$, which is $100\times$ greater than \( \gvec{Q}_{\rm model} \). This setup aims to analyze the system's behavior under significant model-process mismatch.
% \begin{figure}[t]
%     \centering
%     \includegraphics[width=\linewidth]{figs/VsTimeResults/CV_CA_0/TrainPVA_positionX_MATLAB.pdf} 
%     \caption{Single run: \ac{cv} model, Position X estimation Vs. time}
%     \label{fig:CV_CA_0 Position X SR}
% \end{figure}
%
%%%%%%%%%%%%%%%%%%%%%%%%%%%%%%%%%%%
%%% Position Estimation vs Time %%%
%%%%%%%%%%%%%%%%%%%%%%%%%%%%%%%%%%%
%
\begin{figure}[t]
\centering
\includegraphics[width=\linewidth]{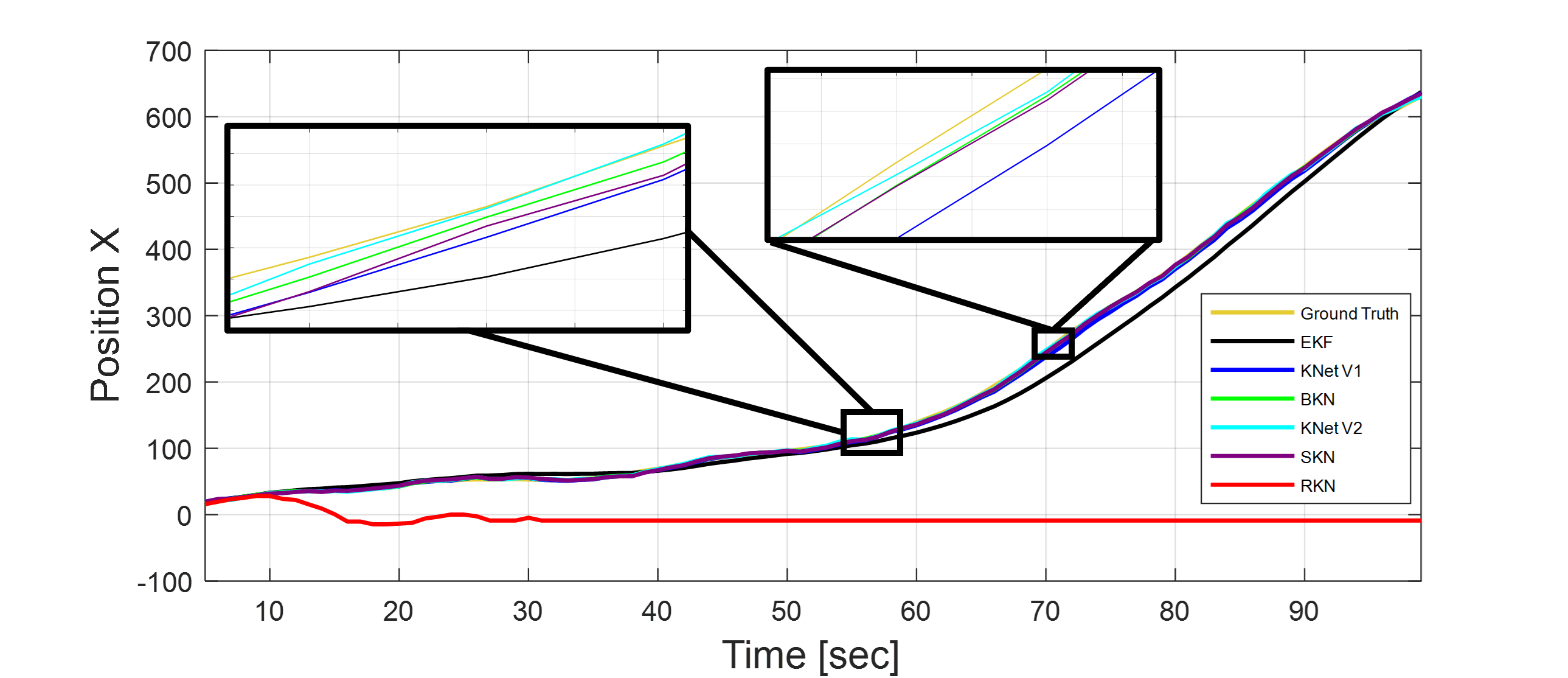} 
\caption{x-axis estimate in a single run, \ac{cv} \ac{ss} model.}
\label{fig:CV_CA_0 Position X SR}
\end{figure}
Fig.~\ref{fig:CV_CA_0 Position X SR} shows the predicted position x track of each method compared to the ground truth trajectory. This figure is a qualitative measure that shows the ability to track the trajectory. As can be seen, the model agnostic \ac{rkn} failed to track the state along time, whereas all the model aware methods provide a reasonable state estimation. 

% \begin{figure}[t]
%     \centering
%     \includegraphics[width=\linewidth]{figs/VsTimeResults/CV_CA_0/MC_TrainPVA_positionX_mc_MATLAB.png}
%     \caption{\ac{cv} \ac{ss}: Monte Carlo: Empirical and Predicted Covariance Vs. time}
%     \label{fig:CV_CA_0 Position X MC}
% \end{figure}
%
\begin{figure*}[t]
%\centering
%
\begin{subfigure}{0.24\textwidth}
\centering
\includegraphics[width=\linewidth]{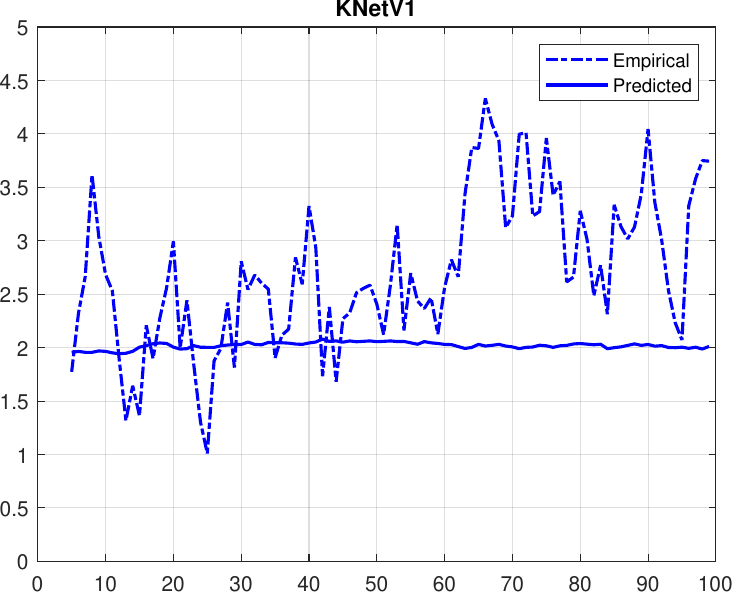}
\caption{KalmanNet V1}
\label{fig:subfig1}
\end{subfigure}
\begin{subfigure}{0.24\textwidth}
\centering
\includegraphics[width=\linewidth]{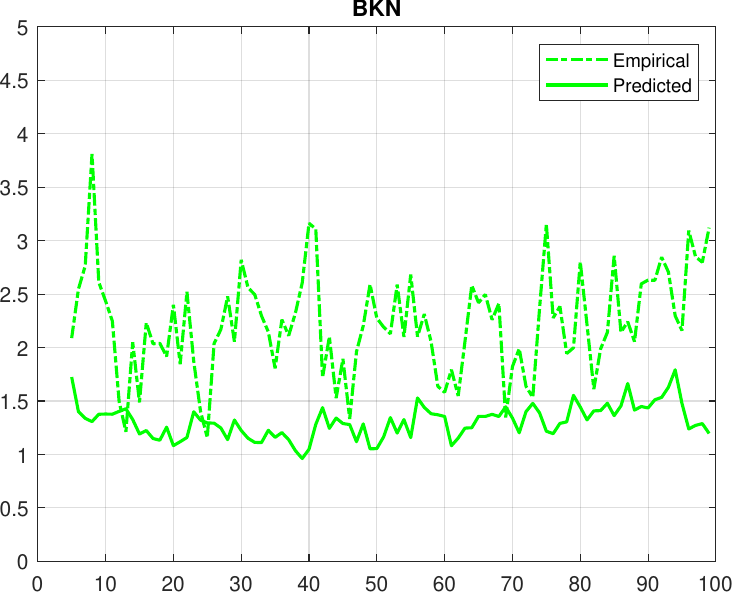}
\caption{Bayesian KalmanNet}
\label{fig:subfig2}
\end{subfigure}
\begin{subfigure}{0.24\textwidth}
\centering
\includegraphics[width=\linewidth]{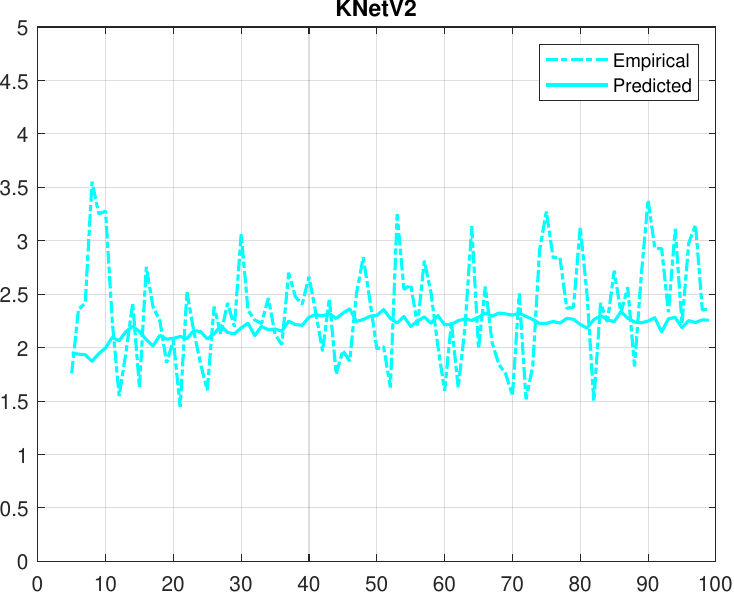}
\caption{KalmanNet V2}
\label{fig:subfig4}
\end{subfigure}
\begin{subfigure}{0.24\textwidth}
\centering
\includegraphics[width=\linewidth]{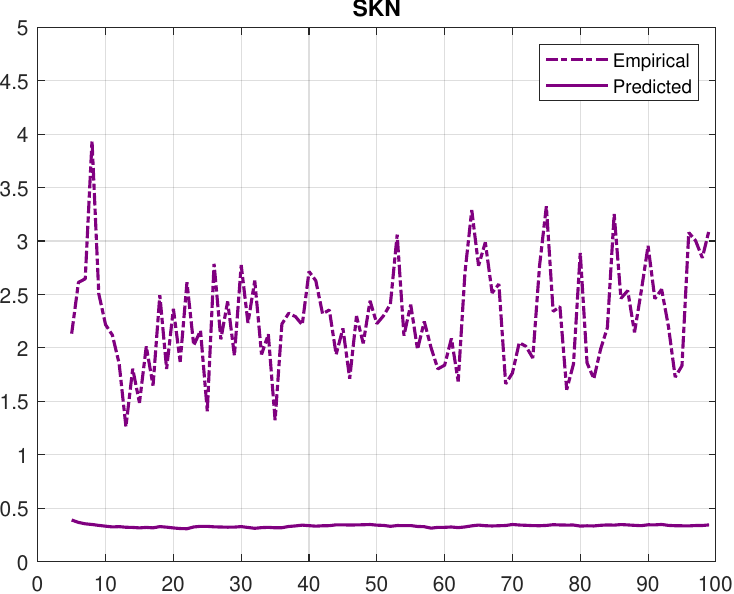}
\caption{Split-KalmanNet}
\label{fig:subfig5}
\end{subfigure}
\caption{The \ac{eec} and \ac{apec} (as described in~\eqref{Average Predicted Covariance} and~\eqref{Empirical Error Covariance}) Vs. time for the model-based deep learning methods.}
\label{fig:CV_CA_0 Position X MC MBDL}
\end{figure*}
%
%%%%%%%%%%%
%%% DNN %%%
%%%%%%%%%%%
%
\begin{figure}
\begin{subfigure}{0.24\textwidth}
%\centering
\includegraphics[width=\linewidth]{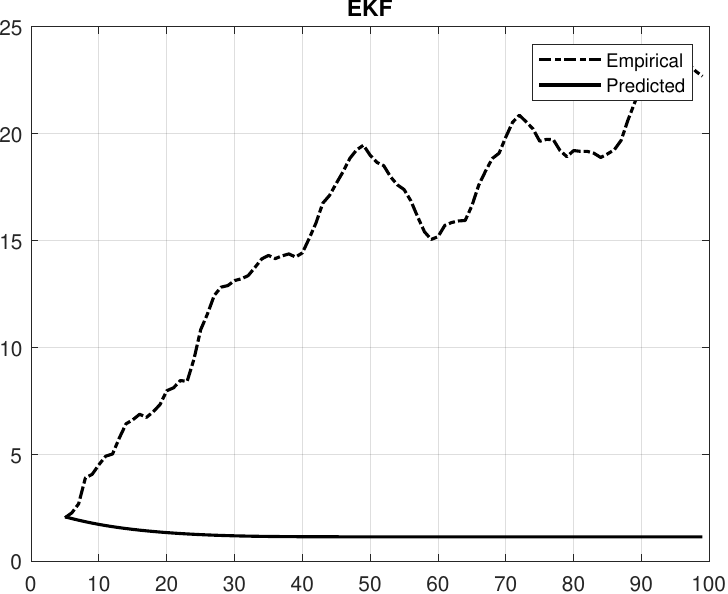}
\caption{Extended Kalman Filter}
\label{fig:subfig3}
\end{subfigure}
\begin{subfigure}{0.24\textwidth}
%\centering
\includegraphics[width=\linewidth]{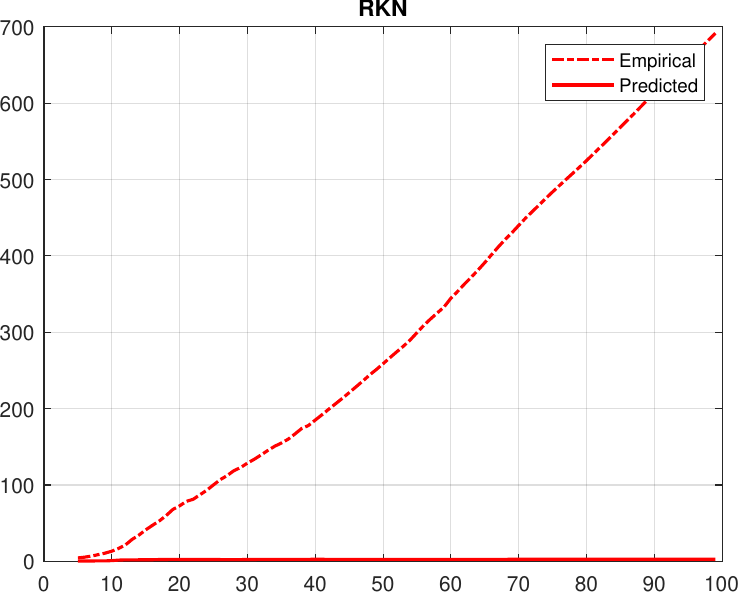}
\caption{Recurrent Kalman Network}
\label{fig:subfig6}
\end{subfigure}
\caption{The \ac{eec} and \ac{apec} (as described in~\ref{Average Predicted Covariance} and~\ref{Empirical Error Covariance}) for the model-based \ac{ekf} and model-agnostic \ac{rkn}.}
\label{fig:CV_CA_0 Position X MC non - MBDL}
\end{figure}

To compare the different methods' performance in prediction the uncertainty, Figs.~\ref{fig:CV_CA_0 Position X MC MBDL} and~\ref{fig:CV_CA_0 Position X MC non - MBDL} show x-axis position  \ac{apec} compared to its \ac{eec}. These metrics are achieved by performing a  simulation with \(N=100\) trajectories, and using \eqref{Average Predicted Covariance} and \eqref{Empirical Error Covariance} to calculate \ac{apec} and \ac{eec} respectively.  As can be seen, the model-based deep learning methods achieve better results than the model-based EKF and the model-agnostic RKN.

% When examining Figure~\ref{fig:CV Mismatch R ANEES} it is clear that for all the tested cases of a \ac{cv} navigation \ac{ss}, the \ac{bkn} architecture has resulted as the most reliable uncertainty predictor.

% 
% %%%%%%%%%%%%%%%%%%%%%
% %%% CV Mismatch R %%%
% %%%%%%%%%%%%%%%%%%%%%
% %
% \begin{figure*}
% \begin{subfigure}{1\columnwidth}
% \centering
% \includegraphics[width=\linewidth]{figs/Results_Full_Paper/CV_CA_5_9_MSE.pdf}
% %\vspace{-0.7cm}
% \caption{\ac{mse}}
% \label{fig:CV Mismatch R MSE}
% %\vspace{-0.3cm}
% \end{subfigure}
% %
% \begin{subfigure}{1\columnwidth}
% \centering
% \includegraphics[width=\linewidth]{figs/Results_Full_Paper/CV_CA_5_9_ANEES.pdf}
% %\vspace{-0.7cm}
% \caption{\ac{anees}}
% \label{fig:CV Mismatch R ANEES}
% %\vspace{-0.3cm}
% \end{subfigure}  
% \caption{\ac{cv} \ac{ss}:mismatched $\gvec{R}$}
% \end{figure*}

\begin{figure*}

\begin{subfigure}{1\columnwidth}
\includegraphics[width=\linewidth]{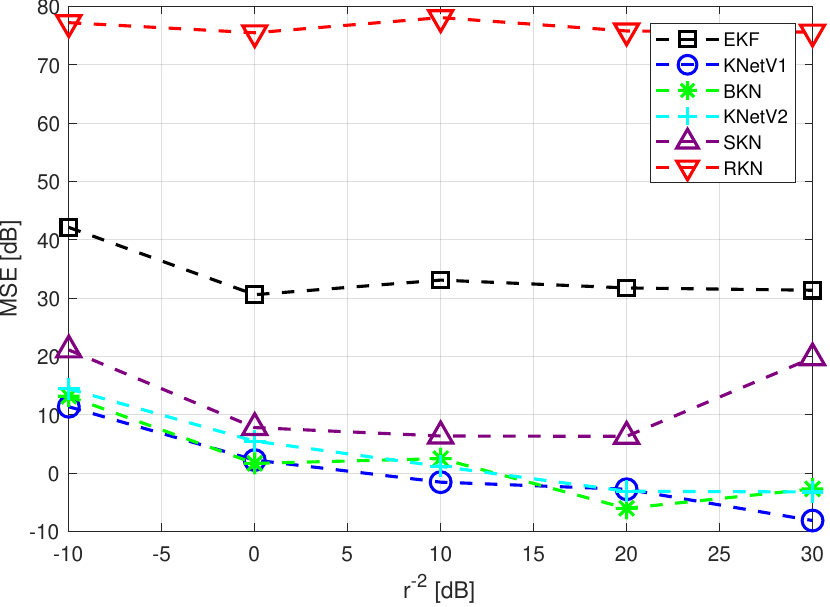}
%\vspace{-0.7cm}
\caption{\ac{mse} for mismatched model, the model based deep learning methods consistently provide the most accurate state estimates compared to the model-free and model dependent methods.
}
\label{fig:CV Mismatch Model MSE}
%\vspace{-0.3cm}
\end{subfigure}
\begin{subfigure}{1\columnwidth}
\includegraphics[width=\linewidth]{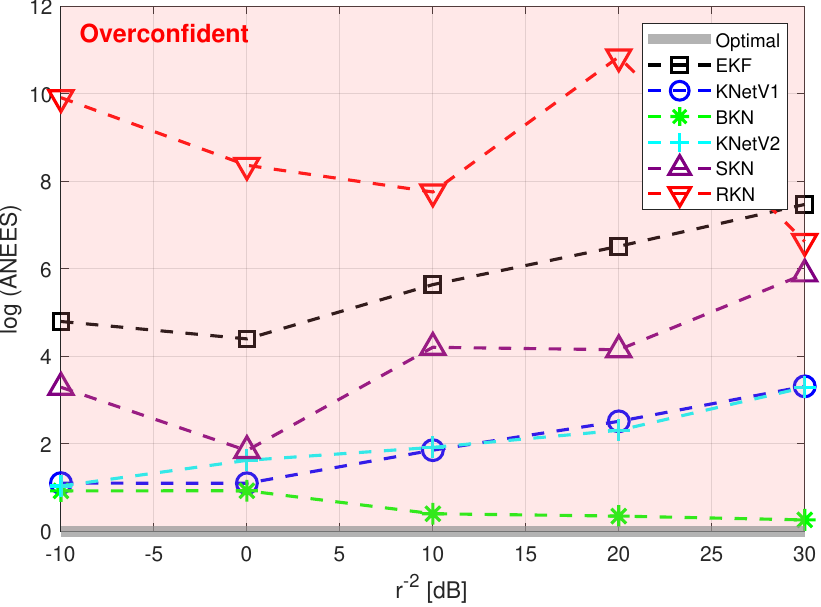}
%\vspace{-0.7cm}
\caption{\ac{anees} for mismatched model, The frequentist KNets have managed to maintain an accurate error covariance estimations, although the \acl{bkn} achieved the most credible error covariance estimation.
}
\label{fig:CV Mismatch Model ANEES}
%\vspace{-0.3cm}
\end{subfigure}
\caption{\ac{cv} \ac{ss} model - mismatched model}
\label{fig:CV Mismatch}
\end{figure*}

The results under model mismatch due to sampling mismatch are reported in Fig.~\ref{fig:CV Mismatch}. Fig.~\ref{fig:CV Mismatch Model ANEES} shows clearly that \acl{bkn} has supplied the most reliable error covariance estimation when given a mismatched model. The model based deep learning frequentist methods also results in a reliable error covariance prediction, whereas the \ac{ekf} fails to extract the uncertainty, outperforming only the model agnostic \ac{rkn}. 
The results for the state estimation when given a mismatched model are very similar to the results in the error covariance estimation. Fig.~\ref{fig:CV Mismatch Model MSE} shows  that the \acl{bkn} has supplied the most reliable state estimation, the frequentist methods  resulted in a reasonable state prediction, whereas the \ac{ekf} and \ac{rkn}  failed to estimate an accurate state estimation.

%
%

%
%%%%%%%%%%%%%%%%%%%%%%%%%
%%% CV Mismatch Model %%%
%%%%%%%%%%%%%%%%%%%%%%%%%
%

%%%%%%%%%%%%%%%%%%%%%%%%%%%%%%%%
%%%%%%%% Inference Time %%%%%%%%
%%%%%%%%%%%%%%%%%%%%%%%%%%%%%%%%
\subsection{Inference Time Comparison}
\label{ssec:results_Inf}

The results presented so far evaluate the considered methods in terms of their accuracy and uncertainty estimation performance. As Bayesian deep learning induces excessive latency in inference compared with frequentist models due to its inherent ensembling, we next evaluate inference time. The results  in Table~\ref{tbl:Runtime} report the inference latency of the different filtering methods across both a linear (canonical) and non-linear (pendulum) scenario. The platform on which all algorithms were tested and runtimes were computed is an Intel(R) Core(TM) i7-10510U CPU @ 1.80GHz 2.30 GHz. \textcolor{NewColor}{For \ac{bkn}, The latency values reflect the total execution time required to process all $J$ realizations sequentially, i.e., the actual inference latency experienced in a serial execution setting.} In the canonical \ac{ss} model, the \ac{ekf} demonstrates the fastest performance of all model aware methods, with the lowest average inference time of $5.279$ milliseconds. The \ac{skn}, KNet V1, and V2 methods exhibit comparable speeds, with marginal differences around $5.3$ milliseconds. These results suggest that these methods maintain efficient performance close to the traditional \ac{ekf}, making them suitable for real-time applications where speed is critical.

The \ac{rkn} achieves an average inference time of just $0.194$ milliseconds. This presumably stems from the efficiency of \ac{rnn} software implementations in PyTorch.  %makes it a compelling choice for scenarios that demand extremely fast processing. On the other hand, the 
\acl{bkn} requires substantially more time, with an inference time of $104.88$ milliseconds for the canonical model. This reflects the computational cost associated with more advanced uncertainty quantification and Bayesian learning techniques. While these methods enhance credibility and robustness, they introduce a significant computational burden, making them less suited for time-sensitive tasks.

In the pendulum scenario, the inference times increase across all methods due to the model’s non-linearity. The \ac{ekf} inference time rises to $13.842$ milliseconds, which is still faster than  \acl{bkn} at $112.461$ milliseconds. Notably, the KalmanNet versions do not have available inference times for the pendulum case, as the measurement matrix $\gvec{H}$ in that \ac{ss} is not full column rank, preventing the calculation of the error covariance. The \ac{skn} retains its relatively low inference time of $13.906$ milliseconds, maintaining its appeal for applications requiring both computational efficiency and accuracy.

%Remarkably, the \ac{rkn} method also demonstrates excellent performance in the non-linear scenario, with an inference time of 0.198 milliseconds. This makes it the fastest method in both the linear and non-linear cases, highlighting its potential for real-time applications.

Overall, while the \acl{bkn} offers substantial benefits in terms of credibility and robustness, these come at the cost of longer inference times. For real-time applications, a parallelization could be used to solve this issue, as the different inferences in the \acl{bkn} are independent and thus could be computed in parallel. \textcolor{NewColor}{While there is no consistent algorithm that achieves the best performance in both accuracy and uncertainty throughout the various settings considered in our numerical study, it is clearly  demonstrated that incorporating domain knowledge notably facilitates uncertainty extraction. Moreover, it is systematically shown that  \acl{bkn} enables \ac{dnn}-aided \acp{kf} to both extract the state and predict the error accurately in nonlinear, possibly mismatched \ac{ss} models, and that its reliability is maintained across all the considered settings.}

\begin{table}[]
\centering
\caption{Average Inference Time [msec].}
\label{tbl:Runtime}
\vspace{0.2cm}
\begin{adjustbox}{width=\columnwidth} 
\begin{tabular}{|c|c|c|c|c|c|c|c|}
\hline
 {\bf{Type}} & {\bf{Model}}  &   {\bf{EKF}}  &   {\bf{KNet V1}}  &   {\bf{KNet V2}}  &   {\bf{SKN}}  &   {\bf{RKN}}  &   {\bf{BKN}}    \\ \hline
 {Linear}    &  {Canonical}  & {5.279}  &   {5.303}  &   {5.279}  &   {5.315}  &   {0.194}\cellcolor[HTML]{AAFDB4}  &   {104.88}\cellcolor[HTML]{FF9595}    \\ \hline % JobID 1013851
 {Non-Linear}   &  {Pendulum}  &{13.842}  &   {N/A}  &   {N/A}  &   {13.906}  &   {0.198}\cellcolor[HTML]{AAFDB4}  &   {112.461}\cellcolor[HTML]{FF9595}    \\ \hline  % JobID 1013954
\end{tabular}
\end{adjustbox}
\vspace{-0.3cm}
\end{table}

\section{Conclusions}
\label{sec:conclusion}
In this work, we studied  uncertainty extraction  in \ac{dnn}-aided \acp{kf}. We identified three techniques based on different indicative features, and proposed two learning schemes that encourage learning to predict the error. Moreover, we have suggested a new method termed \acl{bkn} incorporating Bayesian deep learning to a \ac{dnn} augmented \ac{kf}. Our  numerical results show that incorporating  domain knowledge notably facilitates uncertainty extraction, and that  \acl{bkn} allows \ac{dnn}-aided \acp{kf} to both extract the state and predict the error accurately in non-linear possibly mismatched \ac{ss} models.

%
%%%%%%%%%%%%%%%%%%%
%%% bibliography %%%
%%%%%%%%%%%%%%%%%%%
%
\bibliographystyle{IEEEtran} 
\bibliography{IEEEabrv,ref}

%To add to the final version
\begin{IEEEbiography}[{\includegraphics[width=1 in,height=1.25 in,clip,keepaspectratio]{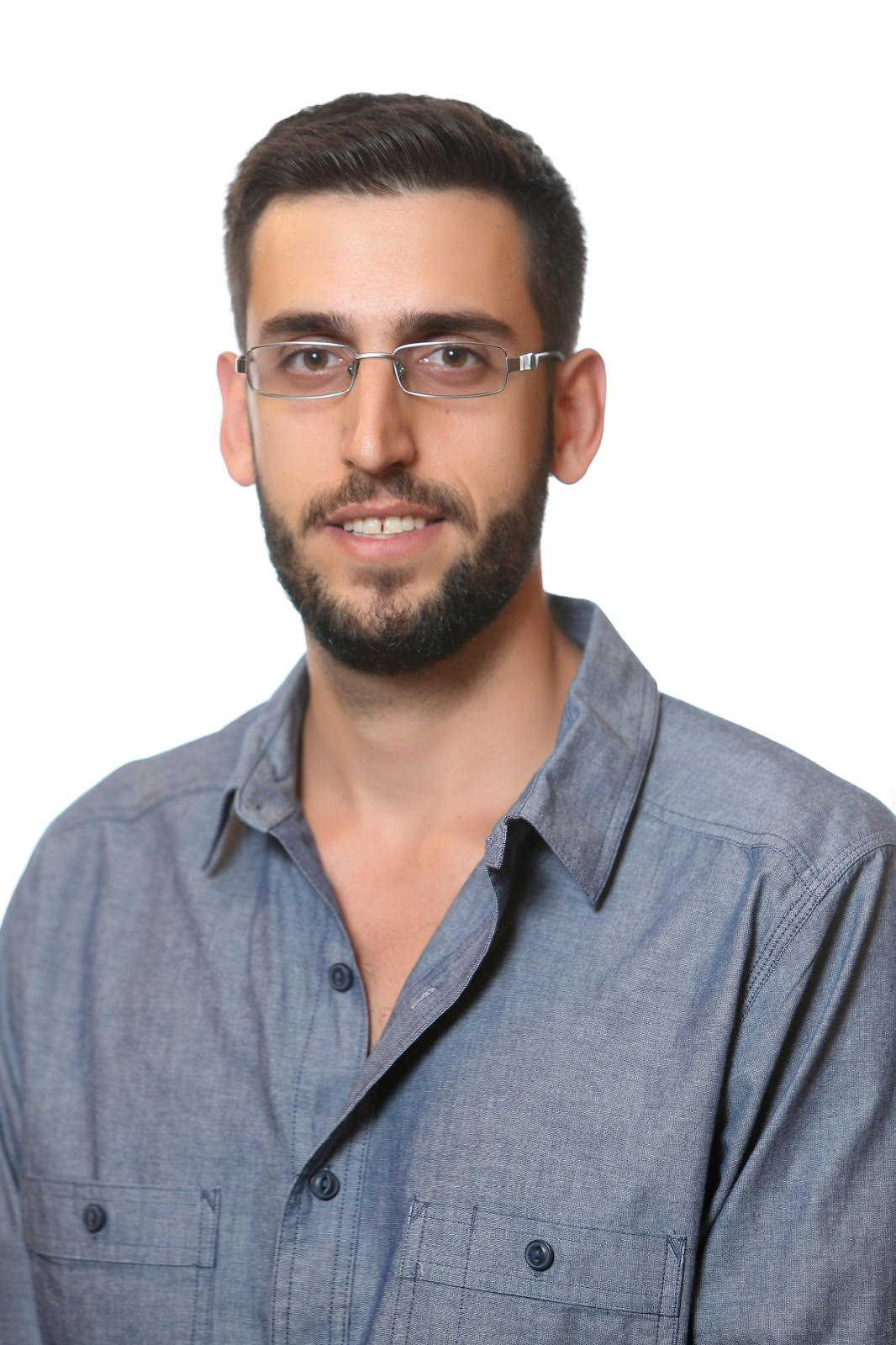}}] {Yehonatan Dahan}
 received his B.Sc. degrees in Mechanical engineering and Materials engineering (double degree with distinction) in 2023, and his M.Sc. degree in Electrical and computer engineering in 2025, all from Ben-Gurion University of the Negev, Israel. His research under the supervision of Dr. Nir Shlezinger include model-based deep learning, Bayesian filtering, state estimation, and their applications in navigation and situational awareness.
\end{IEEEbiography}

\vspace{-0.2cm}

\begin{IEEEbiography} [{\includegraphics[width=1in,height=1.25in,clip,keepaspectratio]{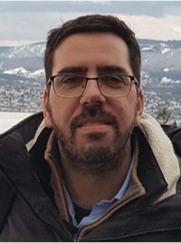}}] {Guy Revach}   is a researcher with a proven industry track record as an innovator and system engineer. He received his B.Sc. (cum laude) and M.Sc. degrees in 2008 and 2017, respectively, from the Andrew and Erna Viterbi Department of Electrical \& Computer Engineering at the Technion – Israel Institute of Technology in Haifa. He completed his master’s thesis under the supervision of Prof. Nahum Shimkin on planning for cooperative multi-agents. Since 2019, he has been a Ph.D. candidate at the Institute for Signal and Information Processing (ISI) at ETH Zürich, Switzerland, supervised by Prof. Dr. Hans-Andrea Loeliger. His main research focuses on the intersection of machine learning with signal processing, explicitly combining sound theoretical principles from classical signal processing with state-of-the-art machine learning algorithms for tracking and detection problems. Before joining ETH Zürich, he worked in the Israeli wireless communication industry for over ten years, initially as a real-time embedded software engineer and later as a software manager. He was the leading innovator behind state-of-the-art, software-defined radio (SDR) for wireless communication, which was game-changing and groundbreaking in size, weight, and power. As a system engineer, he defined significant aspects of SDR requirements and architecture, including hardware, software, network, cyber defense, signal processing, data analysis, and control algorithms.
\end{IEEEbiography} 

\vspace{-0.2cm}

\begin{IEEEbiography}[{\includegraphics[width=1in,height=1.25in, clip,keepaspectratio]{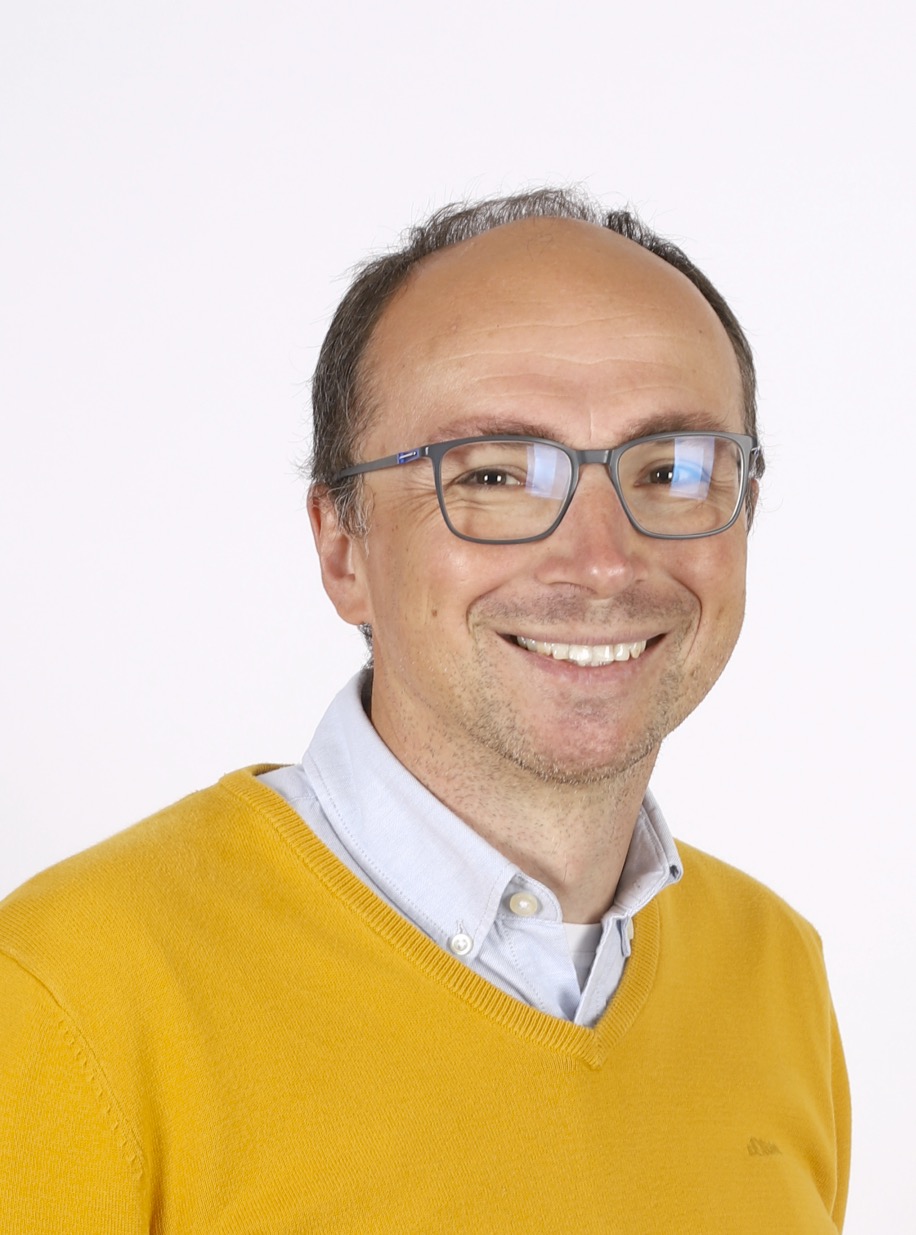}}] {Jindrich Dunik} is an associate professor at the Department of Cybernetics, University of West Bohemia, and a senior scientist with Honeywell International, Aerospace Advanced Technology Europe. He focuses on state estimation, system identification, inertial and satellite-based navigation systems, and integrity monitoring methods. He is a co-author of more than 80 technical papers (both journal and conference) and patents. He is a member of the IEEE AESS Navigation System Panel and serves as Associate Editor of IEEE Transactions on Aerospace and Electronic Systems. He received the Honeywell Technology Achievement Awards, the Werner von Siemens Excellence Award, and the IEEE AESS Harry Rowe Mimno Award.
\end{IEEEbiography}

\vspace{-0.2cm}

\begin{IEEEbiography}[{\includegraphics[width=1in,height=1.25in,clip,keepaspectratio]{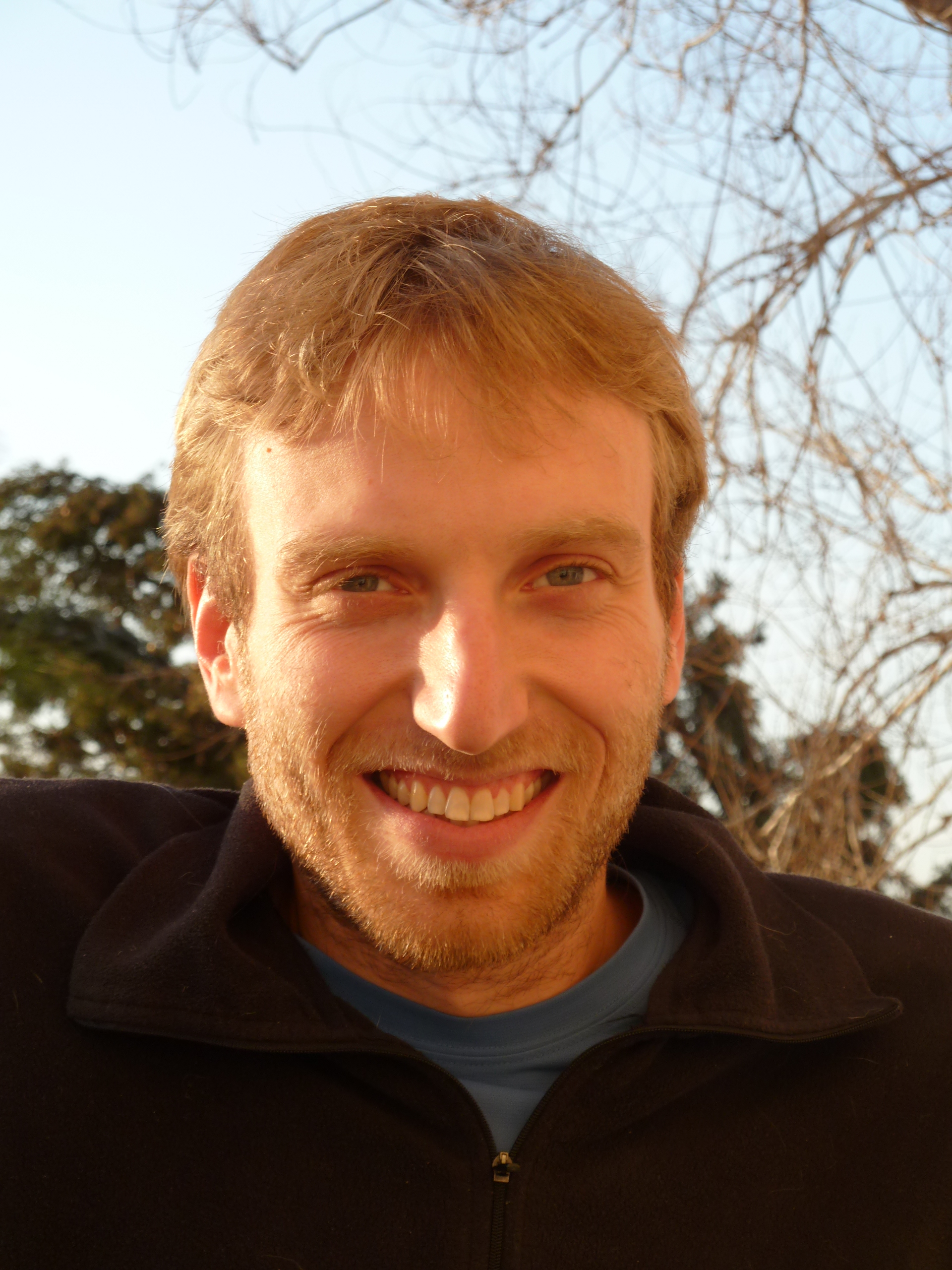}}]{Nir Shlezinger} (M’17-SM'23) is an assistant professor in the School of Electrical and Computer Engineering in Ben-Gurion University, Israel. He received his B.Sc., M.Sc., and Ph.D. degrees in 2011, 2013, and 2017, respectively, from Ben-Gurion University, Israel.
From 2017 to 2019 he was a postdoctoral researcher in the Technion, and from 2019 to 2020 he was a postdoctoral researcher in Weizmann Institute of Science. He was awarded the FGS prize for outstanding postdoctoral research achievements, the 2024 Krill award for young researchers, the Toronto early career award, and the 2024 IEEE Communications Society Fred W. Ellersick prize. 
His research interests include signal processing,  machine learning, and communications.
\end{IEEEbiography}

\end{document}